\documentclass[prd,aps,superscriptaddress,twocolumn,nofootinbib]{revtex4-1}

\usepackage{graphicx} 
\usepackage{amsmath} 
\usepackage{amssymb} 
\usepackage[caption=false]{subfig} 
\usepackage{hyperref} 
\usepackage{physics} 
\usepackage[capitalize]{cleveref} 
\usepackage{bm}
\usepackage{tikz}
\usepackage{tensor}

\usepackage{mathrsfs}

\newcommand{\cP}{\mathcal{P}}%

\newcommand{\deltac}{\delta_\mathrm{c}}
\DeclareMathOperator{\erfc}{erfc}
\DeclareMathOperator{\e}{e}

\graphicspath{{figures/}}

\begin{document}

\title{Excursion-set for Primordial Black Holes I: white noise and moving barrier}

\author{Pierre Auclair}
\email{pierre.auclair@iap.fr}
\affiliation{Sorbonne Université, CNRS UMR 7095, Institut d'Astrophysique de Paris, 98 bis bd Arago, 75014 Paris, France}

\author{Baptiste Blachier}
\email{baptiste.blachier@uclouvain.be}
\affiliation{Cosmology, Universe and Relativity at Louvain (CURL), Institute of Mathematics and Physics, University of Louvain, 2 Chemin du Cyclotron, 1348 Louvain-la-Neuve, Belgium}

\author{Vincent Vennin}
\email{vincent.vennin@ens.fr}
\affiliation{Laboratoire de Physique de l’Ecole Normale Supérieure, ENS, CNRS, Université PSL, Sorbonne Université, Université Paris Cité, F-75005 Paris, France}

\begin{abstract}

In the excursion-set formalism, the mass distribution of primordial black holes (PBHs) is derived from the first-passage time of a random walk describing the density contrast as the coarse-graining scale varies. We address two recent concerns that have been raised about this approach. 
First, it was argued that the random walks are subject to colored (i.e.\ correlated over time) noise, making the first-passage-time problem cumbersome. We show that this arises from an incorrect separation of drift and noise when sampling on the Hubble-crossing surface: if Fourier modes are uncorrelated, the noise is strictly white. Moreover, sampling along the Hubble-crossing surface precludes using the density dispersion as a time variable, explaining some pathologies. Sampling instead on a synchronous surface removes both issues. This requires solving a first-passage-time problem with a moving barrier, for which we provide an efficient numerical framework. 
Second, it was suggested that cloud-in-cloud (i.e.\ that large black holes may engulf smaller ones) is irrelevant for PBHs and that the excursion set is therefore not needed. While valid for widely separated scales, this statement fails for broad power spectra with enhanced continua of modes. We further show that Press–Schechter estimates neglecting boundary evolution can break down even without cloud-in-cloud effects.
\end{abstract}

\maketitle

\section{Introduction}

Large primordial density fluctuations in the early Universe may lead to the formation of primordial black holes (PBHs)~\cite{Carr:1974nx,Carr:1975qj}. The possible existence of such objects has encountered a renewed interest as they could account for (a fraction or all of) dark matter~\cite{Chapline:1975ojl}, act as seeds for early structure formation~\cite{Meszaros:1975ef} or for the presence of supermassive black holes in galactic nuclei~\cite{Duechting:2004dk, Kawasaki:2012kn}. From an observational standpoint, they could further explain some of the black-hole mergers detected by gravitational-wave experiments~\cite{Chiba:2017rvs, LISACosmologyWorkingGroup:2023njw}, the stochastic background of gravitational waves showcased by recent pulsar-timing-array surveys~\cite{EPTA:2023xxk, NANOGrav:2023hvm}, and account for reported microlensing events~\cite{Niikura:2017zjd, Niikura:2019kqi,Sugiyama:2021xqg}.

Various aspects of the PBH hypothesis are under current scrutiny, ranging from formation scenario, calculations of their abundance, mass function and clustering properties, evaporation, merger rates and late-time evolution, see for instance Refs.~\cite{Escriva:2022duf, LISACosmologyWorkingGroup:2023njw}. 
In the present paper, we focus on PBHs originating from the Hubble reentry of large \emph{Gaussian} density fluctuations. Several approaches have been used in this context to predict some of the aforementioned PBHs properties, e.g.\ Press-Schechter formalism~\cite{Press:1973iz}, peak theory~\cite{Bardeen:1985tr, Young:2020xmk} or $N$-body simulations~\cite{Inman:2019wvr,Tkachev:2020uin,Trashorras:2020mwn}.

Among them, the excursion-set framework~\cite{Peacock:1990zz, Bower:1991kf, Bond:1990iw} has the ability to describe hierarchical structure formation, i.e.\ the fact that different structures form at different scales and may encapsulate one another. 
In practice, the density field is coarse grained over a given distance $R$ around a reference location, and when decreasing $R$ it evolves stochastically as new Fourier modes contribute to its realization. 
Structures form when the field overtakes certain formation thresholds, with the largest structure of a given type corresponding to the first ``time'' this occurs. 
Here, ``time'' simply refers to any decreasing labeling of $R$ and does not need to be related to
physical time, as we shall further elaborate on below. 
At the technical level, this amounts to solving a first-passage time problem, where the barrier corresponds to the PBH formation threshold, to which the nature and the statistics of PBHs are exponentially sensitive~\cite{Carr:1975qj, Shibata:1999zs, Musco:2018rwt, Escriva:2019phb, Escriva:2021aeh}. 
This automatically accounts for the ``cloud-in-cloud'' problem~\cite{Jedamzik:1994nr}, namely the fact that large PBHs may form in regions already containing smaller PBHs, in which case only the largest PBH needs to be listed in the final census. 

Recently, concerns have been raised~\cite{Kushwaha:2025zpz} regarding the application of the excursion-set program to PBHs, which allegedly requires solving stochastic processes with colored noises -- a technical complication that most studies avoid -- and may even yield negative mass fractions (a result that is manifestly unphysical) if not handled properly. 
We will demonstrate that all these difficulties arise from sampling the excursion set across the Hubble-crossing hypersurface, and that all such pathologies disappear naturally when a synchronous hypersurface is rather used. This comes at the expense of making the barrier ``time'' (i.e.\ $R$) dependent, but we will show how first-passage time problems with moving barriers can be readily solved.
Such ideas were already discussed in Ref.~\cite{Auclair:2020csm} in the context of the preheating instability; in the present article we generalize and systematize the approach.

On a different note, the excursion-set approach has been used to argue that the cloud-in-cloud phenomenon is irrelevant for PBHs~\cite{MoradinezhadDizgah:2019wjf, DeLuca:2020ioi}, and that its occurrence is suppressed by the abundance of the black holes themselves. 

We will explain why this conclusion holds in the limit where black holes form at well-separated scales, but generically fails otherwise, in particular if they arise from broad power spectra. We will even exhibit situations where the standard Press-Schechter approach breaks down and predicts negative mass functions, including for narrow power spectra. 
By construction, the excursion-set formalism is free from such an anomaly.

The rest of this paper is organized as follows. After reviewing the excursion-set formalism in \cref{sec:General} and discussing the role played by the spacelike hypersurface along which sampling is performed, in \cref{sec:FPT} we show how first-passage time problems with moving barriers can be solved efficiently using a particular set of Volterra integral equations.  
In \cref{sec:Applications}, we apply our framework to PBH formation scenarios with top-hat and (double) log-normal power spectra for illustration. We emphasize the importance of properly accounting for the time dependence of the barrier, and show that although qualitative pictures can sometimes be obtained with constant-barrier approximations -- which are analytically tractable~\cite{Auclair:2024jwj} --
quantitative results based on neglecting the motion of barriers~\cite{Dizon:2025siw, Kameli:2025qzp} might need to be tempered. 
In \cref{sec:results}, we display the associated PBH mass functions, and we discuss the importance of cloud-in-cloud.
Finally, in \cref{sec:Conclusion}, we discuss some limitations of the excursion-set framework and possible extensions of this work. 
In particular, an accompanying paper will focus on applying the procedure of \cref{sec:General,sec:FPT} to the generic formulas of Ref.~\cite{Auclair:2024jwj} for PBH initial clustering.

\section{Excursion-set formalism}
\label{sec:General}

\begin{figure}[t]
    \centering
    \includegraphics[width=\linewidth]{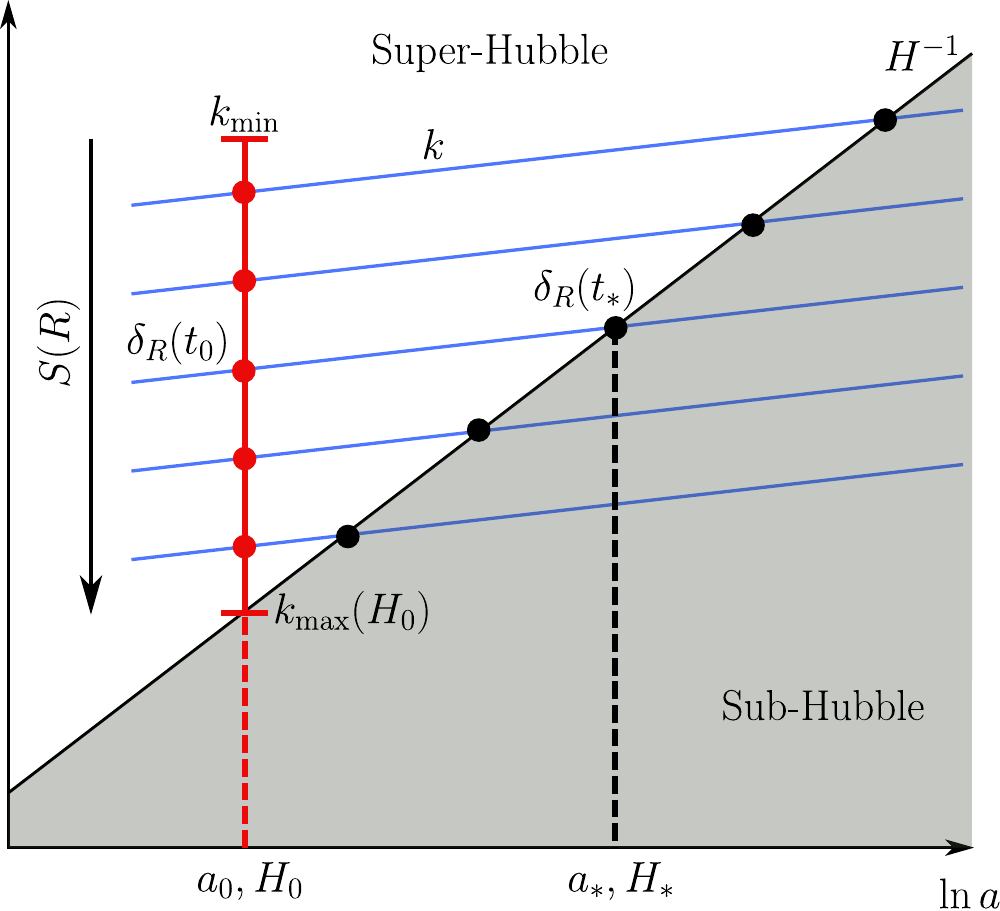}
    \caption{Schematic view of the two choices of sampling of the excursion-set. PBHs form when the value of the coarse-grained density contrast $\delta_R$ is above a certain constant threshold $\deltac$, at the time $t_*$ where the smoothing scale $R$ crosses in the Hubble radius $H^{-1}$ (black line). Different scales $R$ are thus associated to comoving wave numbers $k$ (blue lines) reentering the Hubble radius at different times. 
    Only super-Hubble Fourier modes contribute to $\delta_R$, $\delta_R(t_*)$ and $\delta_R(t_0)$ can be readily related, but working along fixed-time hypersurfaces (red vertical line) has two advantages: (i) it leads to Langevin processes with vanishing drift and (ii) it allows one to relabel $R$ by $S$, leading to Langevin processes with normalized white noises.}
    \label{fig:whitening}
\end{figure}

\subsection{Coarse graining and random walks}

Given a density-contrast field $\delta(\vb{x}, t)$ and its Fourier transform $\delta_{\vb{k}}(t)$, we want to estimate the probability to form gravitationally bound structures, the distribution of their sizes and their hierarchical arrangement.
In the context of the excursion-set framework, we define the coarse-grained density perturbation over a spherical region of radius $R$ about a point $\vb{x}$ evaluated at a certain time $t$ as
\begin{equation}
    \delta_{R}(\vb{x},t) = \int \frac{\dd[3]{\vb{k}}}{(2\pi)^{3/2}} \delta_{\vb{k}}(t) \widetilde{W}\qty(\frac{kR}{a}) \e^{-i\vb{k} \cdot \vb{x}}
    \label{eq:delta_Fourier}
\end{equation}
where $\widetilde{W}$ is a window function in Fourier space selecting the subset of wave numbers $k \lesssim a/R$. In agreement with cosmological observations~\cite{Planck:2019kim}, we assume that the primordial density field has Gaussian statistics with reduced power spectrum $\cP_\delta$ defined as
\begin{equation}
    \langle \delta_{\vb{k}} \delta_{\vb{k}'} \rangle(t) = \frac{2\pi^2}{k^3} \cP_\delta (k, t) \delta_{\mathrm{D}} \left(\vb{k} + \vb{k}'\right)\ ,
\end{equation}
where $\delta_{\mathrm{D}}$ denotes the Dirac distribution. As a consequence, the coarse-grained field $\delta_R$ is also Gaussian, with variance $S(R, t) \equiv \ev{\delta_{R}^{2}(\vb{x},t)}$ given by
\begin{equation}
    S(R, t) =  \int_{0}^{\infty} \cP_{\delta}(k, t) \widetilde{W}^2 \left(\frac{kR}{a} \right) \dd{\ln k} \ .
    \label{eq:S_and_R}
\end{equation}
Let us now study how the coarse-grained field changes when $R$ is varied. 
This can be done along different spacelike hypersurfaces, and in full generality those can be parametrized by a function $t(R)$.
As a consequence,  $\delta_R(t)=\delta_R[t(R)]$ becomes a function of $R$ only (and we omit the argument $\vb{x}$ hereafter since it is held fixed). 
One obtains 
\begin{equation}
    \label{eq:langevin}
    \dv{\delta_R }{R} = t'(R) \dot{\delta}_R + \xi(R)
\end{equation}
where $\dot{\delta}_R$ is the coarse-grained time derivative of the density field, and 
\begin{equation}
    \xi(R) =  \int \frac{\dd[3]{\vb{k}}}{(2\pi)^{3/2}} \delta_{\vb{k}}(t)  \frac{\partial}{\partial R}\left\lbrace\frac{kR}{a[t(R)]}\right\rbrace    \widetilde{W}'\qty(\frac{kR}{a}) \e^{-i\vb{k} \cdot \vb{x}}\, .
\end{equation}
In \cref{eq:langevin}, the first term is driven by the evolution of the mode functions $\delta_{\vb{k}}[t(R)]$ along the sampled hypersurface, and plays the role of a ``drift''. The second term models the contribution of the Fourier modes crossing $R$ and joining the coarse-grained field as $R$ decreases. As such, \cref{eq:langevin} can be interpreted as describing a random walk: as $R$ decreases, new modes contribute to the integral of \cref{eq:delta_Fourier}, and since each of these modes takes a random realization, $\xi$ plays the role of a ``noise'' with centered Gaussian statistics. Since the field $\delta_R$ is statistically homogeneous and isotropic, the statistical properties of such random walk are independent of $\vb{x}$, hence by taking ensemble averages over realizations of \cref{eq:langevin} one effectively evaluates spatial averages.

For this description to be consistent, modes need to cross the coarse-graining scale outward, which implies that $R/a[t(R)]$ decreases as $R$ decreases. This yields a constraint on the sampling hypersurface, namely that 
\begin{equation}
    \label{eq:cond:1}
    R t'(R) H[t(R)]<1\, .
\end{equation}
Note also that an important simplification arises from working with a top-hat window function in Fourier space, $\widetilde{W}(kR/a)=1$ if $kR/a<1$ and $0$ otherwise. In this case, the two-point function of the noise reads as
\begin{multline}
    \label{eq:2pt:xi}
    \left\langle \xi(R) \xi(R') \right\rangle = 
    \left (\dv{ }{R} \left\{ \ln \frac{R}{a[t(R)]} \right\}\right)^2 \mathcal{P}_\delta\left[k=\frac{a}{R},t(R)\right] \\ 
    \times \delta_{\mathrm{D}}\left(R-R'\right) ,
\end{multline}
namely the noise is white (i.e.\ uncorrelated over ``time'' $R$), and \cref{eq:langevin} is a Langevin equation that describes a Markovian process.\footnote{Non-sharp window functions would make the noise colored, as recently discussed in Ref.~\cite{Saito:2025sny} in the PBH context, see also Refs.~\cite{Paranjape:2011wa, Musso:2011ck, Musso:2013pha, Nikakhtar:2018qqg} in the context of large-scales structures.}

\subsection{Sampling along the Hubble-crossing hypersurface}
\label{sec:sampling:Hubble:surface}

A first possible choice for the sampling hypersurface is the Hubble-crossing surface, i.e.\ $R(t)=H^{-1}(t)$, which implicitly defines the function $t(R)$ and leads to $t'(R)=1/\epsilon_1$ where $\epsilon_1=-\dot{H}/H^2$ is the first Hubble-flow parameter. Then, the condition~\eqref{eq:cond:1} reduces to $\epsilon_1>1$, i.e.\ the expansion needs to decelerate, which is indeed the case once inflation has ended. One advantage of this choice is that the PBH collapse threshold $\deltac$ is often expressed in terms of the comoving density contrast~\cite{Young:2014ana, Auclair:2020csm} at Hubble crossing time.\footnote{Numerical simulations~\cite{Musco:2018rwt, Germani:2018jgr, Escriva:2019phb, Musco:2020jjb} have shown that the critical threshold depends sensitively on the details of the density profile around the overdensity peak and may vary, see \cref{sec:Conclusion}. Note that if non-Gaussianities are involved, the criterion may also depend on the very nature of the fluctuations triggering the PBH collapse~\cite{Shimada:2024eec, Inui:2024fgk, Escriva:2025eqc, Escriva:2025rja}.} This implies that PBH formation corresponds to crossing a fixed ``barrier'' at $\delta_R=\deltac$. 

In this case, since only super-Hubble modes contribute to $\delta_R$, and given that the comoving density contrast behaves as $\delta_{\vb{k}}(t) \propto (aH)^{-2}\propto a^{1+3w} $ on super-Hubble scales, one has $\dot{\delta}_R = (1+3w) H \delta_R$. Here, for simplicity we assume that PBHs form during an epoch where the equation-of-state parameter $w$ is constant. The Langevin equation~\eqref{eq:langevin} thus becomes 
\begin{equation}
    \label{eq:langevin:Hubble:crossing}
    \dv{\delta_R }{R} = \frac{2}{3}\frac{1+3w}{1+w}\delta_R+ \xi(R),
\end{equation}
where the noise obeys \cref{eq:2pt:xi} with $\dv{ }{R} \{ \ln R/{a[t(R)]} \} = \frac{1}{R}\frac{1+3w}{3(1+w)}$.

One may identify two problems about this approach.
First, it is reported in Ref.~\cite{Kushwaha:2025zpz} that the noise is non-white, which seemingly contradicts the above discussion. This is because, in Ref.~\cite{Kushwaha:2025zpz} the Langevin equation is written as $\dv*{\delta_R}{R}=\tilde{\xi}$, where $\tilde{\xi}$ corresponds to the whole right-hand side of \cref{eq:langevin:Hubble:crossing}. 
As such, $\tilde{\xi}$ is correlated over time since $\delta_R$ is. However, when properly separating contributions from the drift and the noise, the noise is always white, if the window function is sharp in Fourier space. There is a way to get rid of the drift and make  $\tilde{\xi}$ white that we discuss in the next subsection (see footnote~\ref{footnote:rescaling:drift:removal}) but let us stress that $\xi$ is by no means colored and \cref{eq:langevin:Hubble:crossing} always describes a Markovian process.

Second, it has been raised that negative mass functions can be obtained in some regimes for instance in Ref.~\cite{Sureda:2020vgi}. 
Mass functions will be further discussed in \cref{subsec:mass_func}, where we will explain why their computation is usually performed by relabeling the ``time'' variable $R$ by $S$ in the Langevin equation. For this change of variable to be well defined, the function $S[R,t(R)]$ needs to be monotonically decreasing. With top-hat Fourier-space window functions, \cref{eq:S_and_R} leads to
\begin{equation}
    R \dv{S }{R} = 
    4\frac{\epsilon_1-1}{\epsilon_1}S
    - \left(1-\frac{1}{\epsilon_1}\right)\mathcal{P}_\delta\left[k=aH,t(R)\right], 
\end{equation}
where we have used that $\delta_{\vb{k}}(t) \propto a^{1+3w} $ on super-Hubble scales. We have also expressed the equation-of-state parameter $w=2\epsilon_1/3-1$ in terms of the first Hubble-flow parameter to make it clear that, under the condition~\eqref{eq:cond:1}, i.e.\ $\epsilon_1>1$, the first term is positive while the second term is negative. As a consequence, $\dv*{S}{R}$ is not necessarily negative in this setup: at scales where $\mathcal{P}_\delta$ assumes small values, the first term in the above may dominate over the second one and ruin the monotonic mapping between $R$ and $S$.
This is why Ref.~\cite{Kushwaha:2025zpz} introduces a new variable $\tau = R^{-1}$.

\subsection{Sampling along a synchronous hypersurface}
\label{subsec:synchronous}

\begin{figure}[t]
    \centering
    \includegraphics[width=.49\textwidth]{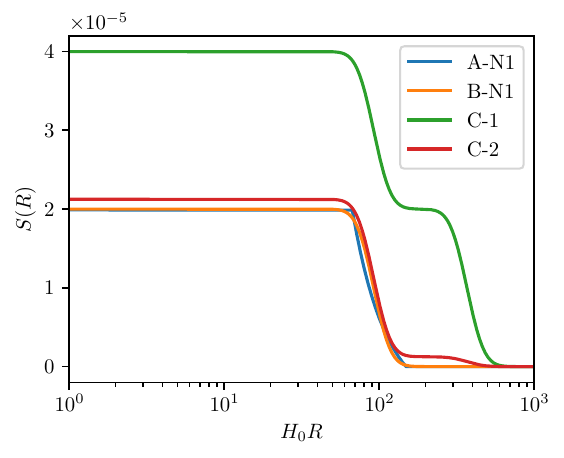}
    \caption{Variance $S$ against coarse-graining scale $R$ computed for some of the benchmark scenarios listed in \cref{sec:Applications}, see \cref{tab:params}.}
    \label{fig:variance}
\end{figure}

Another choice of sampling hypersurface corresponds to one with constant $t=t_0$. In this case, $t'(R)=0$ so \cref{eq:cond:1} is immediately satisfied and the drift in \cref{eq:langevin} vanishes altogether. The Langevin equation becomes
\begin{equation}
    \label{eq:langevin:synchronous}
    \dv{\delta_R }{R} =  \xi(R),
\end{equation}
where the noise obeys \cref{eq:2pt:xi} with $\dv{ }{R} \{ \ln R/{a[t(R)]} \} = R^{-1}$. The ``time'' relabeling $R\to S$ is also always valid, since
\begin{equation}
\label{eq:S'(R)}
    R \dv{S }{R} = 
    - \mathcal{P}_\delta\left[k=a(t_0)/R,t_0\right]<0
\end{equation}
in that case. This allows one to rewrite \cref{eq:langevin:synchronous} in the convenient form 
\begin{equation}
    \label{eq:langevin:synchronous:S}
    \dv{\delta_R }{S} =  \xi(S),
    \quad\quad \left\langle \xi(S) \xi(S')\right\rangle = \delta_{\mathrm{D}}(S-S')
\end{equation}
where $\xi(S)$ is a white Gaussian noise, with vanishing mean and unit variance.
 
The price to pay is that the PBH formation threshold needs to be expressed at a fixed time rather than at Hubble crossing. At a given $R$, if $t_*$ denotes the time when $R=H^{-1}$ and provided $t_0<t_*$, one has
\begin{equation}
    \delta_R(t_*)=\left[\frac{a(t_*)H(t_*)}{a(t_0)H(t_0)}\right]^{-2}\delta_R(t_0)
\end{equation}
since both $\delta_R(t_*)$ and $\delta_R(t_0)$ are made of super-Hubble Fourier modes. In  the following, we denote with a subscript $0$ (resp. $*$) quantities evaluated at $t_0$ (resp. $t_*$).

As a consequence, the PBH formation threshold $\delta_{\mathrm{c}}$, that applies to $\delta_R(t_*)$, can be redshifted to the time $t_0$ where it applies to $\delta_R(t_0)$, leading to 
\begin{equation}
    \label{eq:deltac:rescaling}
    \delta_{\mathrm{c}} \left(R \right)= \left(H_0 R\right)^{-\frac{2}{3}\frac{1+3w}{1+w}} \deltac \, .
\end{equation}
In practice, this leads to a ``time'' (i.e.\ $R$)-dependent barrier.
\footnote{Note that the rescaling $\delta_R\to R^{-\frac{2}{3}\frac{1+3w}{1+w}} \delta_R$ of \cref{eq:deltac:rescaling} is precisely the one that makes the drift in \cref{eq:langevin:Hubble:crossing} disappear, which is consistent.\label{footnote:rescaling:drift:removal}} 
The situation is summarized in \cref{fig:whitening}.  Along the $t=t_0$ hypersurface, starting from $S=0$ (or equivalently $R = \infty$) the first ``time'' when $\delta_R$ encounters the collapse threshold $\delta_{\mathrm{c}}(R )$ provides the size of the largest  PBH surrounding the point $\vb{x}$. The statistics of this first-passage time is thus related to the size distribution of black holes, which we further study below.

Before closing this section, let us finally note that, owing to the simplicity of \cref{eq:langevin:synchronous:S,eq:deltac:rescaling}, all the dependence on initial conditions is contained in the relationship between $S$ and $R$, that is required to turn \cref{eq:deltac:rescaling} into a barrier function $\delta_{\mathrm{c}}(S)$. It is convenient to express \cref{eq:S_and_R} in terms of the power spectrum of curvature perturbations $\zeta$, given that those are conserved at super-Hubble scales and can thus be directly evaluated at the end of inflation.  At super-Hubble scales, $\zeta$ is related to the density contrast in the comoving gauge according to
\begin{equation}
    \delta_{\vb{k}}(t) \simeq - \frac{2 (1 + w)}{5 + 3w} \left( \frac{k}{aH} \right)^2 \zeta_{\vb{k}} \ ,
    \label{eq:comoving_density}
\end{equation}
hence 
\begin{equation}
    \label{eq:S(R)}
    S(R)= \left[\frac{2 (1 + w)}{5 + 3w}\right]^2 \int_0^{(H_0 R)^{-1}} \!\!\! u^4 \mathcal{P}_\zeta(k=a_0 H_0 u) \dd\ln u\, .
\end{equation}
One can thus check that, when $R\to \infty$, $S\to0$. There is however a maximal value to $S$: from \cref{fig:whitening}, it is clear that only curvature perturbations at scales $k < k_{\mathrm{max}} = a_0 H_0$  can be resolved. 
This implies that $R>R_{\mathrm{min}}=H_0^{-1}$, which in turns leads to a maximal value for $S$, denoted $S_{\mathrm{max}}$ below. 
In practice, this does not limit the application of the excursion-set formalism since it suffices to match $t_0$ with the end of inflation (or any time prior to the reentry of the peak of the power spectrum, as discussed in \cref{sec:UV}) in order to include all relevant scales. 
However, it does imply that the first-passage-time problem must be solved in the presence of an upper bound on the ``time'' parameter.

\begin{figure}[t]
    \centering
    \includegraphics[width=.49\textwidth]{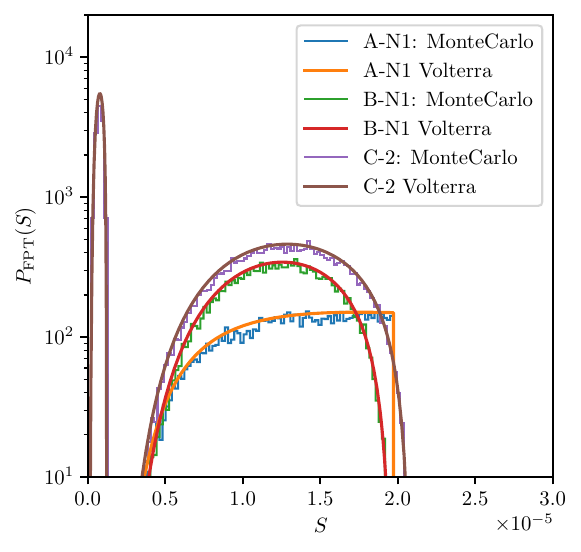}
    \caption{First-passage time obtained through solving the Volterra equation numerically against MonteCarlo simulation with $10^7$ trajectories and $1000$ time steps. The various models used are identified in \cref{tab:broad,tab:lognorm,tab:double} and will be discussed in \cref{sec:Applications}.
    Note that despite the relatively large number of trajectories, the noise in the MonteCarlo simulation is still large. This is because only a small subset of trajectories end up crossing the barrier.
    Of course, one could use importance sampling to obtain better convergence. 
    }
    \label{fig:MCvsMatrix}
\end{figure}

\section{First-passage time problem}
\label{sec:FPT}

\subsection{Moving barrier and Volterra equations}
\label{sec:Volterra}

\begin{figure*}[t]
    \centering
    \subfloat[Model A-N1]{\includegraphics[width=.31\textwidth]{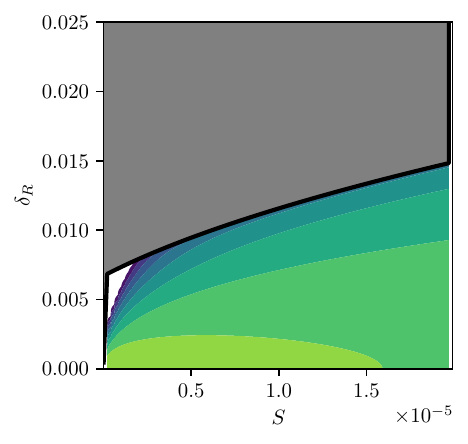}\label{fig:kernel-a}}
    \subfloat[Model B-N1]{\includegraphics[width=.31\textwidth]{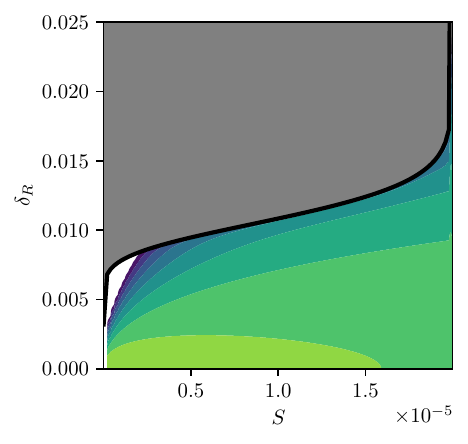}\label{fig:kernel-b}}
    \subfloat[Model C-1]{\includegraphics[width=.38\textwidth]{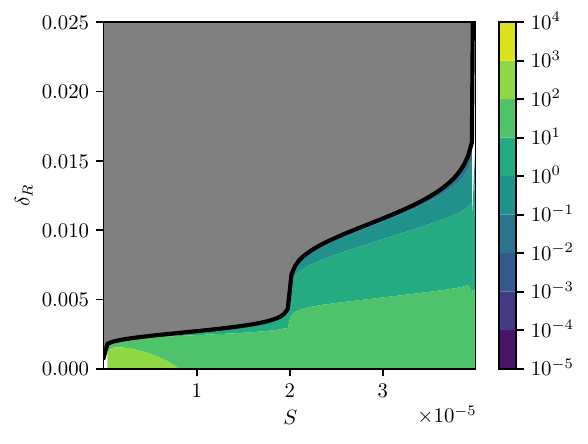}\label{fig:kernel-c}}
    \caption{Probability kernel computed through \cref{eq:proba_transition} for different benchmark models. 
    The solid black line corresponds to the location of the time-dependent boundary $\deltac(S)$.
    }
    \label{fig:kernel}
\end{figure*}

We now explain how one can solve the first-passage time problem we are left with, namely a Brownian motion with a moving barrier. This problem, well known in probability theory, only admits very few analytical results and therefore numerics must be used. The availability of numerically inexpensive methods~\cite{Tuckwell_Wan_1984, Buonocore1987ANI, Zhang:2005ar, Molini2011, Auclair:2020csm} is the reason why formulating the excursion set on synchronous hypersurfaces is so efficient.

First and foremost, the Brownian motion performed by the density contrast $\delta_R$ for given initial conditions $(\delta_{R,\mathrm{i}}, S_{\mathrm{i}})$ is characterized by its transition probability without enforcing any boundary condition. It is given by the Gaussian distribution
\begin{equation}
    \label{eq:free}
    P_{\mathrm{free}} (\delta_{R}, S \vert \delta_{R,\mathrm{i}}, S_{\mathrm{i}})
    = \frac{e^{- \frac{(\delta_{R} - \delta_{R,\mathrm{i}})^{2}}{2(S-S_{\mathrm{i}})}}}{\sqrt{2\pi(S-S_{\mathrm{i}})}} \ .
\end{equation}
Then, when an absorbing condition is imposed at an arbitrary time-dependent threshold $\deltac(S)$, the first-passage time probability distribution $P_{\mathrm{FPT}}$ obeys the implicit relation~\cite{Buonocore1987ANI, Giorno:1989, Zhang:2005ar, Auclair:2020csm}
\begin{multline}
    P_{\mathrm{FPT}}(S \vert \delta_{R,\mathrm{i}}, S_{\mathrm{i}}) = \left[\frac{\deltac(S) - \delta_{R,\mathrm{i}}}{S-S_{\mathrm{i}}} - 2 \deltac'(S) + K(S)\right] \\ \cross P_{\mathrm{free}}\left[\deltac(S), S\vert \delta_{R,\mathrm{i}}, S_{\mathrm{i}}\right]  + \int_{S_{\mathrm{i}}}^{S} \dd{s} P_{\mathrm{free}} \left[\deltac(S), S\vert\deltac(s), s\right] \\ \cross P_{\mathrm{FPT}}(s \vert \delta_{R,\mathrm{i}}, S_{\mathrm{i}}) \left[2 \deltac'(S) - \frac{\deltac(S) - \deltac(s)}{S-s} - K(S) \right] .
    \label{eq:Volterra}
\end{multline}
This is a Volterra integral equation of the second kind with a generic kernel function $K(S)$. As explained in details in~\cite{Buonocore1987ANI,Auclair:2020csm}, although \cref{eq:Volterra} is satisfied for any kernel function $K(S)$, the kernel can be chosen in order to remove the singularities -- and thus numerical instabilities -- plaguing the integrand of the above equation close to the upper bound, at $s\simeq S$. 
This singularity is removed when choosing $K(S) = \deltac'(S)$.

It should also be noted that, if one sets $K(S) = 0$, the first term in the Volterra equation is nothing but twice the flux of trajectories across the boundary $\delta_{\mathrm{c}}(S)$
\begin{multline}
    P_\mathrm{PS}(S \vert \delta_{R,\mathrm{i}}, S_{\mathrm{i}}) \equiv 
    \left[\frac{\deltac(S) - \delta_{R,\mathrm{i}}}{S-S_{\mathrm{i}}} - 2 \deltac'(S)\right] \\
    \cross P_\mathrm{free}\left[\deltac(S), S\vert \delta_{R,\mathrm{i}}, S_{\mathrm{i}}\right] .
    \label{eq:ps}
\end{multline}
Up to a factor of $2$, we recover the distribution used to compute the Press-Schechter (PS) mass function~\cite{Press:1973iz}, see Ref.~\cite{Auclair:2020csm} for a detailed discussion.

Upon the discretization of the time variables $S = n \Delta s + S_{\mathrm{i}}$ and $s = m\Delta s$ with $n,m$ integers and $\Delta s$ a numerical step\footnote{The discretization scheme is made explicit for a fixed step $\Delta s$, but in practice one can discretize the Volterra equation with any slicing of $S$.}, this allows one to recast \cref{eq:Volterra} as the following linear matrix equation
\begin{equation}
    \label{eq:Volterra:matrix}
    P_{\mathrm{FPT}} = P_\mathrm{PS} + J P_{\mathrm{FPT}} \Delta s + \deltac'(S) P_\mathrm{free}\, .
\end{equation}
In this formula, $P_{\mathrm{FPT}} = P_{\mathrm{FPT}} (n\Delta s)$ is a $n$-dimensional vector, and so is $P_\mathrm{PS}^n$, defined as
\begin{multline}
    P_\mathrm{PS}^n = \left[  \frac{\deltac (n\Delta s + S_{\mathrm{i}}) - \delta_{R,\mathrm{i}}}{n\Delta s} - 2 \deltac' (n\Delta s + S_{\mathrm{i}})\right] \\ \cross P_{\mathrm{free}} \left[ \deltac (n\Delta s + S_{\mathrm{i}}) - \delta_{R,\mathrm{i}}, n\Delta s \right]  .
    \label{eq:x}
\end{multline}
Finally, $J$ is a lower triangular matrix, with vanishing diagonal and
\begin{multline}
    J\indices{^{n}_{m}} = P_{\mathrm{free}} \left[ \deltac (n\Delta s + S_{\mathrm{i}}) - \deltac (m\Delta s + S_{\mathrm{i}}), (n-m) \Delta s \right] \\ \cross \left[\deltac'(n \Delta s + S_{\mathrm{i}}) - \frac{\deltac(n \Delta s + S_{\mathrm{i}}) - \deltac(m \Delta s + S_{\mathrm{i}})}{(n-m) \Delta s} \right]   .
\end{multline}
The matrix equation is then solved according to
\begin{equation}
    P_{\mathrm{FPT}} = (\mathrm{Id} - J \Delta s)^{-1} [P_\mathrm{PS} + \deltac'(S) P_\mathrm{free}]
    \label{eq:matrix-eq}
\end{equation}
where the lower triangular matrix $\mathrm{Id} - J \Delta s$ can be efficiently inverted using dedicated algorithms.\footnote{A triangular matrix can be inverted in $\order{n^2}$ using forward substitution as opposed to $\order{n^3}$ for the simplest algorithm of matrix inversion, $\order{n^{\log_2(7)}} \approx \order{n^{2.8074}}$ for Strassen's algorithm included in BLAS~\cite{strassen1969gaussian,press2007numerical} and $\order{n^{2.371339}}$ for the current best theoretical bound \cite{doi:10.1137/1.9781611978322.63}.}

By comparison with a standard Monte-Carlo simulation, this method proves to be much more efficient in terms of numerical cost and reliability. It can be checked with \cref{fig:MCvsMatrix} where the two methods are compared for several power spectra which will be introduced in \cref{sec:Applications} together with the parameters given in \cref{tab:broad,tab:lognorm,tab:double}. Furthermore, quite remarkably, the sampling $n$ of the Volterra integral equation does not need to be high to ensure a fast-convergence for the first-passage time distribution, see \cref{sec:converge}.

From the knowledge of $P_\mathrm{FPT}$, one can also obtain the transition probability $ P(\delta_R, S \vert \delta_{R,\mathrm{i}}, S_{\mathrm{i}} )$ associated to the random motion starting at $(\delta_{R,\mathrm{i}}, S_{\mathrm{i}})$ when the (moving) barrier is accounted for :
\begin{multline}
    P(\delta_R, S \vert \delta_{R,\mathrm{i}}, S_{\mathrm{i}} ) = P_{\mathrm{free}}(\delta_R, S | \delta_{R,\mathrm{i}}, S_{\mathrm{i}}) \\ - \int_{0}^{S} \dd{s} P_{\mathrm{FPT}}(s | \delta_{R,\mathrm{i}}, S_{\mathrm{i}}) P_{\mathrm{free}} \left[ \delta_R - \deltac(s), S-s | \delta_{R,\mathrm{i}}, S_{\mathrm{i}}   \right] .
    \label{eq:proba_transition}
\end{multline}
We will call this quantity the ``probability kernel'' and we show it for different benchmark models in \cref{fig:kernel}. 
The two quantities given by \cref{eq:Volterra,eq:proba_transition} suffice\footnote{In order to avoid numerical instabilities near the boundary of the integral in \cref{eq:proba_transition}, a trick can be employed to rewrite this equation and ensure high numerical precision, see \cref{sec:num_tips}.} to completely characterize the random motion performed by the density contrast with respect to $S$. In particular, we explain in the next subsection how the PBH mass distribution can be obtained from the knowledge of the barrier-crossing distribution.

\subsection{Mass function}
\label{subsec:mass_func}

For a given (stochastic) realization of $\delta_R$, there are as many substructures as the number of times $\delta_R$ crosses the threshold $\deltac (S)$. However, since PBHs engulf their inner substructure, one only needs to keep track of the largest structure formed. The excursion-set formalism allows us to pick precisely those largest objects, thus avoiding overcounting and by design solving the ``cloud-in-cloud'' problem.

PBHs forming at the scale $R$ have a mass of the order of the Hubble mass at the time $R$ reenters the Hubble radius,\footnote{If the collapsing region already contains a black hole, its density does not necessarily redshift as the background density since black holes behave as nonrelativistic matter. This is why, in principle, the mass estimate~\eqref{eq:M(R)} has to be refined if cloud-in-cloud takes place.}
\begin{equation}
    \label{eq:M(R)}
    M(R)=4\pi \sigma \frac{M_{\mathrm{Pl}}^2}{H\left[t_*\left(R\right)\right]} = M_0 \left(H_0 R\right)\, ,
\end{equation}
where $\sigma \sim 0.1$ is a parameter measuring the efficiency of the collapse, $M_{\mathrm{Pl}}$ is the reduced Planck mass, and $M_0$ is the mass of PBHs forming at time $t_0$. 
The above formula provides a rough estimate only, while more refined methods involve the use of  critical scaling laws~\cite{Choptuik:1992jv, Niemeyer:1997mt, Musco:2008hv, Musco:2012au} or rely on the compaction function~\cite{Shibata:1999zs,Harada:2015yda,Musco:2018rwt}. 
These methods are however more challenging to implement within the excursion-set approach, and we further discuss them in \cref{sec:Conclusion}.

Let $\beta(M)$ denote the PBH mass fraction at formation time, i.e.\ $\beta(M) \dd \ln M$ corresponds to the fraction of the Universe comprised in PBHs of masses $[ M, M + \dd M ]$ when black holes of mass $M$ form. Then, from the excursion-set viewpoint,
\begin{equation}
    \beta(M) \dd \ln M = - P_\mathrm{FPT} (S) \dd{S} \, ,
    \label{eq:M&Pfpt}
\end{equation}
where $S$ and $M$ are related via \cref{eq:S(R),eq:M(R)}. This is why, as mentioned in \cref{sec:sampling:Hubble:surface}, if the function $S(R)$ is not monotonically decreasing, then one may get ``negative'' mass distributions, but this simply signals that $S$ cannot be used to relabel time in the stochastic process in that case. 
The use of a synchronous hypersurface always prevents this issue from occurring.

After PBHs form, they behave as nonrelativistic matter, hence their fractional density scales as $a^{3w}$. If $f(M)$ denotes the PBH mass fraction at some fixed, late reference time $t_{\mathrm{ref}}$, after all PBHs formed, one thus has
\begin{equation}
\label{eq:f(M)}
    f(M) = \beta(M) \left[\frac{a(t_{\mathrm{ref}})}{a_*(M)}\right]^{3w}\!\! = \left(\frac{a_{\mathrm{ref}}}{a_0}\right)^{3w}\!\!\beta(M) \left(\frac{M_0}{M}\right)^{\frac{2w}{1+w}} .
\end{equation}

\section{Applications}
\label{sec:Applications}

The method laid out above was used to study metric preheating in Ref.~\cite{Auclair:2020csm}. Here, we further illustrate its application to the case where black holes form in the radiation era ($w=1/3$) with a formation threshold $\deltac = 1$ in \cref{eq:deltac:rescaling}.
In this context, it has been suggested that power spectra with narrow peaks give rise to nearly monochromatic mass functions~\cite{Niemeyer:1997mt, Yokoyama:1998xd, Green:1999xm} (although some of the assumptions leading to this result might be questioned~\cite{Germani:2023ojx}). 
In practice, most inflationary models producing large fluctuations at small scales come with a peak of finite width in the power spectrum~\cite{Pi:2022zxs, Domenech:2023dxx, Cielo:2024poz, Briaud:2025hra}, and one may expect PBHs to form over a finite range of masses. 
It has also been argued that cloud-in-cloud remains marginal, and suppressed by the PBH abundance itself, even in the presence of broad power spectra \cite{MoradinezhadDizgah:2019wjf, DeLuca:2020ioi}. 
In this work, we wish to reexamine these claims by considering several power spectra motivated by the literature on PBHs.

\begin{table*}[t]
    \centering
    \subfloat[Top-hat spectrum]{
        \begin{tabular}{c | ccc | c}
            Identifier & $\mathcal{A}$ & $k_1$ & $k_2$ & $S_\mathrm{max}$ \\
            \hline \hline
            A-N1 & $2.28$ & $8.19 \times 10^{-2}$ & $1.22 \times 10^{-1}$ & $2 \times 10^{-5}$\\
            A-N2 & $1.14$ & $8.19 \times 10^{-2}$ & $1.22 \times 10^{-1}$ & $10^{-5}$\\
            A-N3 & $0.57$ & $8.19 \times 10^{-2}$ & $1.22 \times 10^{-1}$ & $5 \times 10^{-6}$\\
            A-W1 & $849.1$ & $2.707 \times 10^{-4}$ & $1.478 \times 10^{-2}$ & $2 \times 10^{-6}$ \\
            A-W2 & $84.91$ & $2.707 \times 10^{-4}$ & $1.478 \times 10^{-2}$ & $2 \times 10^{-7}$ \\
            A-W3 & $8.491$ & $2.707 \times 10^{-4}$ & $1.478 \times 10^{-2}$ & $2 \times 10^{-8}$
        \end{tabular}
        \label{tab:broad}
    } %
    \subfloat[Log-normal spectrum]{
        \begin{tabular}{c | ccc | cc}
            Identifier & $\mathcal{A}$ & $k_{\mathrm{p}}$ & $\Delta$ & $k_\mathrm{min}$ & $S_\mathrm{max}$ \\
            \hline \hline
            B-N1 & $9.34 \times 10^{-1}$ & $10^{-1}$ & $10^{-1}$ & $10^{-3}$ & $2 \times 10^{-5}$\\
            B-N2 & $4.67 \times 10^{-1}$ & $10^{-1}$ & $10^{-1}$ & $10^{-3}$ & $10^{-5}$\\
            B-N3 & $2.35 \times 10^{-1}$ & $10^{-1}$ & $10^{-1}$ & $10^{-3}$ & $5 \times 10^{-6}$\\
            B-W1 & $212.2$ & $2 \times 10^{-3}$ & $1$ & $10^{-5}$ & $2 \times 10^{-6}$\\
            B-W2 & $21.22$ & $2 \times 10^{-3}$ & $1$ & $10^{-5}$ & $2 \times 10^{-7}$\\
            B-W3 & $2.122$ & $2 \times 10^{-3}$ & $1$ & $10^{-5}$ & $2 \times 10^{-8}$
        \end{tabular}
        \label{tab:lognorm}
    }\\
    \subfloat[Double log-normal spectrum, based on B-N1]{
        \begin{tabular}{c | ccccc | cc}
            Identifier & $\mathcal{A}_1$ & $\mathcal{A}_2$ & $k_{\mathrm{p}, 1}$ & $k_{\mathrm{p}, 2}$ & $\Delta$ & $k_\mathrm{min}$ & $S_\mathrm{max}$ \\
            \hline \hline
            C-1 & $2^4 \times \mathcal{A}_2$ & $9.34 \times 10^{-1}$ &  $5 \times 10^{-2}$  & $10^{-1}$ & $10^{-1}$ & $10^{-3}$ & $4 \times 10^{-5}$\\
            C-2 & $\mathcal{A}_2$ & $9.34 \times 10^{-1}$ & $5 \times 10^{-2}$ &  $10^{-1}$ & $10^{-1}$ & $10^{-3}$ & $2.125 \times 10^{-5}$\\
        \end{tabular}
        \label{tab:double}
    }
    \caption{Benchmark models, wave numbers are given in units of $a_0 H_0$. A suffix N labels ``narrow peak'' models and a suffix W labels ``wide peak'' models.}
    \label{tab:params}
    
\end{table*}

\begin{figure*}[ht]
    \centering
    \subfloat[Narrow top-hat spectrum]{\includegraphics[width=0.49\textwidth]{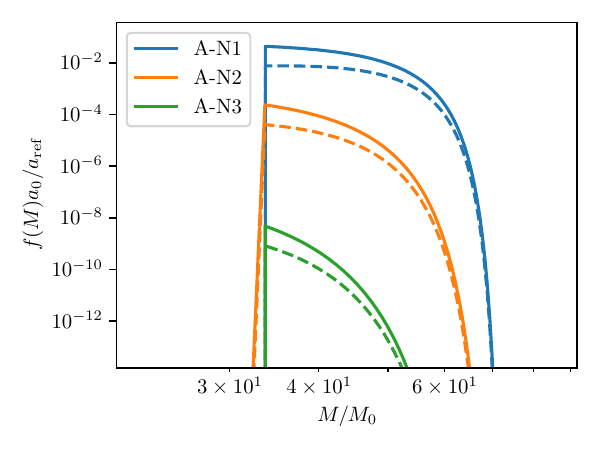}\label{fig:mass_func_fig:an}}%
    \subfloat[Narrow log-normal spectrum]{\includegraphics[width=0.49\textwidth]{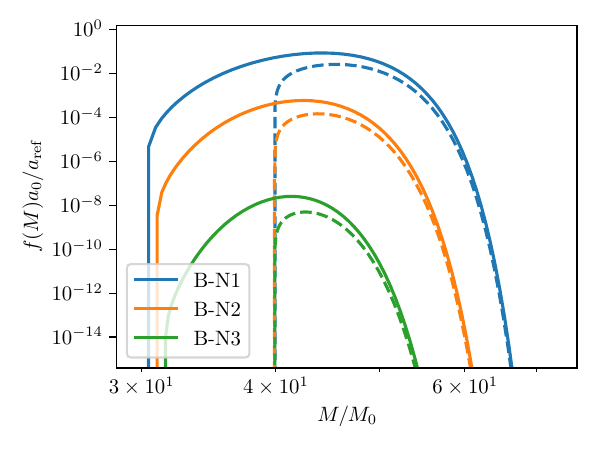}\label{fig:mass_func_fig:bn}} \\ %
    \subfloat[Wide top-hat spectrum]{\includegraphics[width=0.49\textwidth]{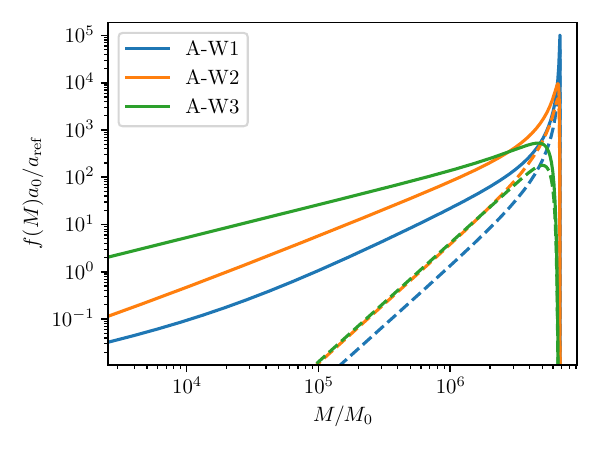}\label{fig:mass_func_fig:aw}}%
    \subfloat[Wide log-normal spectrum]{\includegraphics[width=0.49\textwidth]{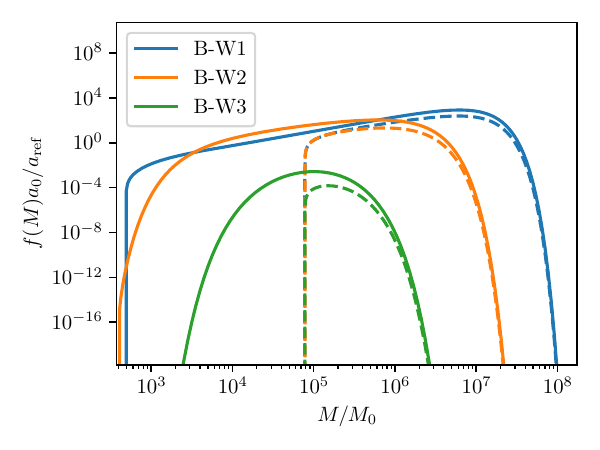}\label{fig:mass_func_fig:bw}} %
    \caption{Mass function for the top-hat power spectrum (left panels) and for the log normal power spectrum (right panels). The dashed lines correspond to the Press-Schechter approach, i.e.\ where cloud-in-cloud is neglected (see main text).
    }
    \label{fig:mass_func_fig}
\end{figure*}

\subsection{Top-hat power spectrum}
\label{subsec:broad}

We first consider an idealized version of a broad power spectrum, frequently used as a benchmark~\cite{Saito:2009jt, MoradinezhadDizgah:2019wjf, Sugiyama:2020roc, DeLuca:2020ioi, DeLuca:2020agl, Fumagalli:2024kgg}, made of a scale-invariant ``peak'' that spreads over the range of modes $[k_1, k_2]$ with an amplitude $\mathcal{A}$:
\begin{equation}
    \label{eq:Pzeta:broad}
    \cP_\zeta (k) = \mathcal{A} \Theta(k - k_1) \Theta (k_2 - k) \ .
\end{equation}
Conveniently, the rescaling~\eqref{eq:S(R)} can be performed analytically,
\begin{equation}
    S(R) = \frac{4}{81} \mathcal{A} \left[ (H_0 R)^{-4} - \left( \frac{k_1}{a_0 H_0} \right)^4 \right] 
    \label{eq:S(R)_broad}
\end{equation}
if $a_0/k_2<R<a_0/k_1$, and $S(R)$ is constant outside that range.
This relation is displayed in \cref{fig:variance} for a benchmark model taken from \cref{tab:broad}. 
Inverting \cref{eq:S(R)_broad} and using \cref{eq:deltac:rescaling}, the time dependence of the barrier is given by
\begin{equation}
\label{eq:deltac(S):tophat}
    \delta_\mathrm{c}(S) = \delta_\mathrm{c} \left[\frac{81}{4 \mathcal{A}} S + \left( \frac{k_1}{a_0 H_0} \right)^4\right]^{1/4}
\end{equation}
for $0\leq S\leq S_{\mathrm{max}}$, where 
\begin{equation}
    S_{\mathrm{max}} = \frac{4}{81}\mathcal{A}\frac{k_2^4-k_1^4}{\left(a_0H_0\right)^4}\, .
\end{equation}
Together with the probability kernel introduced in \cref{eq:proba_transition}, the function $\deltac(S)$ is displayed in \cref{fig:kernel-a}.
\footnote{To ease numerical comparison with other models, a low-amplitude scale-invariant contribution has been added to the power spectrum~\eqref{eq:Pzeta:broad} in all figures, such that the $S(R)$ function remains invertible until $R=a_0/k_{\mathrm{max}}$. The amplitude of that contribution is taken sufficiently low so that it does not affect our quantitative results.}

\subsection{Log-normal peak}
\label{subsec:lognorm}

In order to avoid the (nonphysical) sharp features arising from the edges of a top-hat power spectrum, a smoother version is often considered in the form of a log-normal peak~\cite{Young:2019osy, Pi:2020otn, Germani:2023ojx, Pi:2024ert},
\begin{equation}
    \cP_\zeta (k) = \frac{\mathcal{A}}{\sqrt{2\pi} \Delta} \exp \left[ - \frac{\ln(k/k_{\mathrm{p}})^2}{2\Delta^2}   \right]  .
    \label{eq:Pw_lognorm}
\end{equation}
The quantity $\mathcal{A}$ still controls the amplitude, $k_{\mathrm{p}}$ represents a certain scale at which the power spectrum peaks, whereas $\Delta$ parametrizes the width of the peak. For $\Delta \ll 1$, the above reduces to a Gaussian peak, characteristic of the class of narrow-peak power spectra. Again, the rescaling~\eqref{eq:S(R)} can be performed analytically,
\begin{align}
    S(R) =& \frac{8\mathcal{A}}{81} \left( \frac{k_{\mathrm{p}}}{a_0 H_0} \right)^4 e^{8\Delta^2} \nonumber \\ 
    &\times \Bigg\{ \erf \left[ \frac{1}{\sqrt{2}\Delta} \ln \left(\frac{a_*}{Rk_{\mathrm{p}}} \right) - 2\sqrt{2} \Delta \right]  \nonumber\\ 
    &-  \erf \left[ \frac{1}{\sqrt{2}\Delta} \ln \left(\frac{k_\mathrm{min}}{k_{\mathrm{p}}} \right) - 2\sqrt{2} \Delta \right]\Bigg\}  ,
    \label{eq:S(R)-lognorm}
\end{align}
where $a_*$ depends on $R$ through $a_*(R) = a_0 \sqrt{H_0 R}$. Here we have introduced a minimum scale $k_\mathrm{min}$, so the above expression is valid only for $R < a_*(R) /k_{\mathrm{min}}$, since otherwise $\deltac(S=0)=0$ and $\deltac'(S=0)=\infty$, which cannot be numerically handled.
In \cref{subsec:IR}, we further check that $k_\mathrm{min}$ has been set to sufficiently low values such that it does not affect our physical results.
For the sake of comparison, we select the parameters of this model to match qualitatively the top-hat power spectrum.
In particular, the peak and the width are chosen such that
\begin{equation}
    \label{eq:k12:Delta}
    k_{1,2} = k_{\mathrm{p}} \exp(\pm 2 \Delta)\, ,
\end{equation}
and the amplitude $\mathcal{A}$ is set to obtain the same $S_\mathrm{max}=S(R=a_0/k_{\mathrm{max}})$, see~\cref{tab:broad,tab:lognorm}.

Looking at \cref{fig:variance}, one can check that for narrow peaks (with a suffix N) the function $S(R)$ has a similar shape to the one obtained with a top-hat power spectrum, although smoother. 
The $S$-dependence of the threshold is found by inverting \cref{eq:S(R)-lognorm} numerically, and it is displayed, together with the probability kernel extracted from \cref{eq:proba_transition}, in \cref{fig:kernel-b}. 
The behavior is again similar to (although smoother than) the one obtained with the top-hat power spectrum in \cref{fig:kernel-a}.

However, we have checked that for wider peaks (with a suffix W), the shape of the $S(R)$ function with top-hat and log-normal power spectra differs more significantly, because of the heavy tails of the log-normal distribution. Even if we match the peak and width of the distributions, the two models are no longer qualitatively similar in that regime.

\subsection{Double log-normal spectrum}
\label{subsec:bouble-lognorm}

\begin{figure}
    \centering
    \includegraphics[width=0.49\textwidth]{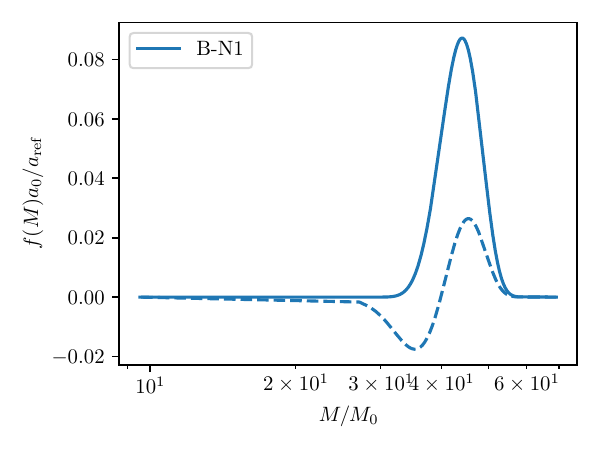}
    \caption{
        Mass function of model B-N1 in linear scale, from both the excursion-set (solid) and the Press-Schechter approximation (dashed). 
        Press-Schechter predicts that the mass function becomes negative in a certain mass range, whereas the excursion-set remains always positive.
        }
    \label{fig:figure_6_fail}
\end{figure}

As a last application, we consider a power spectrum that peaks at two distinct scales. This may seem fine-tuned from an inflationary model-building point of view, but this toy model will prove insightful when it comes to determining under which conditions cloud-in-cloud may or may not be neglected, and ponder the role of mass hierarchies in that discussion. A simple way to achieve a double-peak power spectrum is by adding to 
\cref{eq:Pw_lognorm} a second log-normal 
peak at $k_\mathrm{p,1} < k_\mathrm{p,2}$. The width of the second peak is also taken equal to $\Delta$. If the two peaks are sufficiently well separated, 
we expect the mass function to exhibit 
two distinct, well-resolved maxima. 

The two sets of parameters used for this model are listed in \cref{tab:double}. For the model C-1, they are chosen so that the second peak gives the same contribution to the variance $S_\mathrm{max}$ as the first peak, as can be seen from the green curve in \cref{fig:variance}. 
Conversely, the model C-2 is such that the two peaks have comparable amplitudes, hence the second peak contributes less to the variance $S_\mathrm{max}$.
See the red curve in \cref{fig:variance}.

The $S$-dependence of the threshold along with the probability kernel is represented in \cref{fig:kernel-c} for the model C-1. 
The model C-2 presents a threshold with an analogous shape but the width of the first upward ``step'' is reduced.

\section{On the relevance of the excursion-set approach}
\label{sec:results}

\begin{figure*}[t]
    \subfloat[No additional peak]{\includegraphics[width=.33\textwidth]{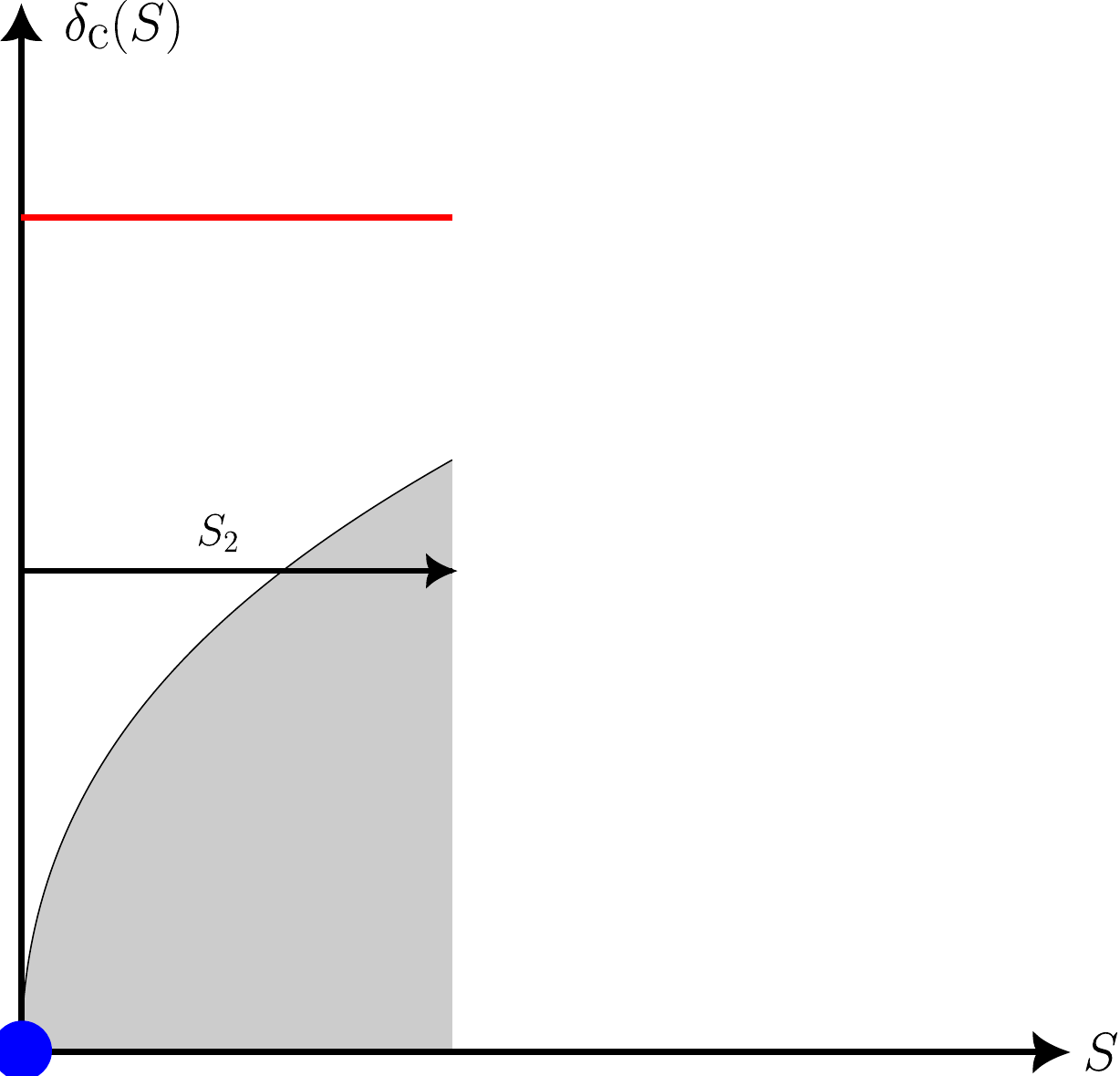}}
    \subfloat[Small peak]{\includegraphics[width=.33\textwidth]{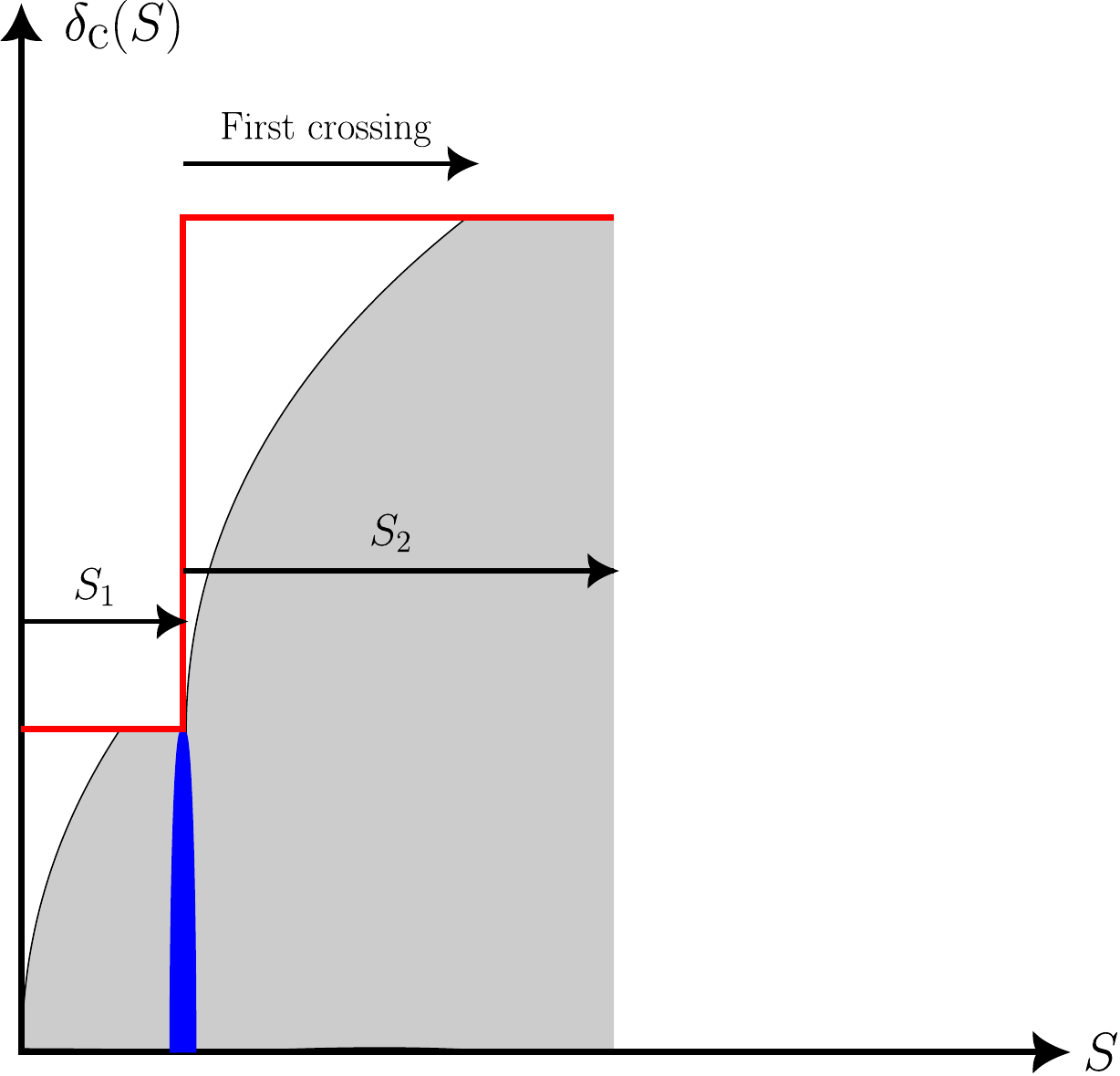}}
    \subfloat[Large peak]{\includegraphics[width=.33\textwidth]{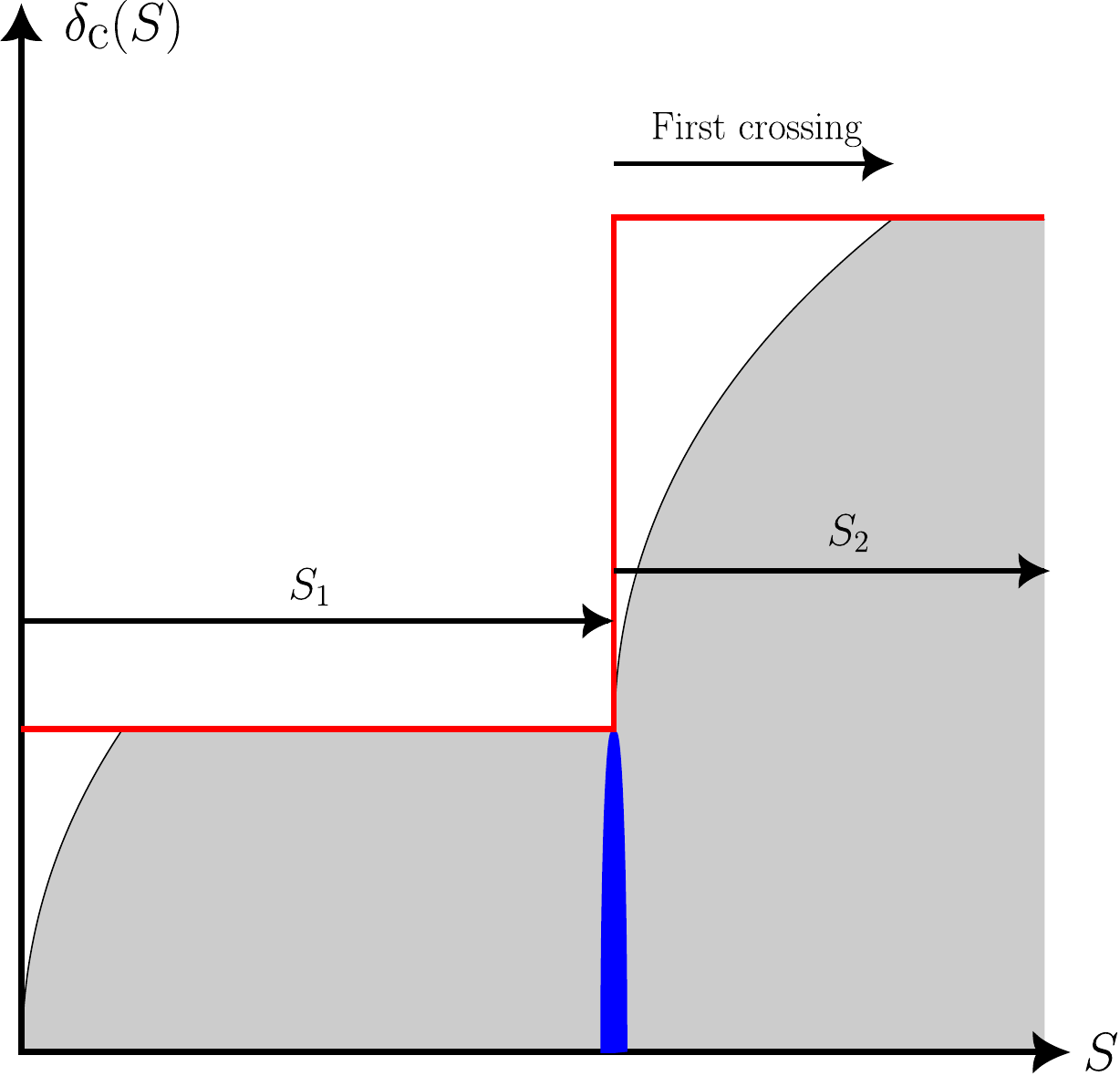}}
    \caption{Schematic view for the effect of a second peak at large scales onto the PBH mass function.
    The gray region shows the typical spread of the Langevin trajectories. 
    The inclusion of a second peak at large scales adds an initial period of diffusion of ``duration'' $S_{1}$, increasing from the leftmost to the rightmost panel.
    At first, the main effect of this initial period of diffusion is to spread initial conditions (blue ellipse), making barrier crossing more likely to occur during the second phase.
    As $S_{1}$ becomes larger, fewer trajectories survive this initial period of diffusion, hence first crossing during the second phase becomes less likely and the abundance of low-mass black holes is reduced.
    This explains the nonmonotonic behavior in \cref{fig:mass_func_double}.
    }
    \label{fig:effect-of-second-peak}
\end{figure*}

For the models listed in \cref{sec:Applications}, the first-passage time distribution can be computed with the method detailed in \cref{sec:Volterra}, from which the mass function can be obtained as explained in \cref{subsec:mass_func}. 
The result is displayed in \cref{fig:mass_func_fig,fig:mass_func_double} for the parameters listed in \cref{tab:broad,tab:lognorm,tab:double}.

\subsection{Broadness of the mass function}

Let us first focus on power spectra with one narrow peak, namely the models A-N1 to A-N3 and B-N1 to B-N3. From the solid lines of \cref{fig:mass_func_fig:an,fig:mass_func_fig:bn}, the top-hat and log-normal power spectra lead to similar mass distributions (both in terms of amplitude and width), which differ only in the far tails. The low-mass cutoff is sharper with the top-hat power spectrum than with the log-normal one, but the latter shall otherwise be seen as a smoothed version of the former. The mass functions are sharply peaked, since the value for $\Delta$ -- or equivalently, for $\ln(k_2/k_1)$, see \cref{eq:k12:Delta} -- we have employed is small, see \cref{tab:broad,tab:lognorm}. This confirms that peaked power spectra give rise to PBH mass functions that are essentially monochromatic (if critical scaling was accounted for, a low-mass heavier tail would nonetheless develop, see \cref{sec:Conclusion} for further discussions).

To the contrary, we find in \cref{fig:mass_func_fig:aw,fig:mass_func_fig:bw} that the PBH distributions for the ``wider peaks'' differ significantly between the top-hat and the log-normal model. This is a direct consequence of the fact that a log-normal distribution with $\Delta = 1$ does not fall off rapidly both in the UV and the IR, and therefore cannot be matched accurately to a top-hat distribution. Nonetheless, we find that for both power-spectrum shapes, the PBH distribution spans several orders of magnitude, up to $5$ for B-W1, when the width of the power spectrum is only of order one ($\Delta = 1$). Crucially, the amplitude of the power spectrum plays a crucial role in determining the shape and the width of the PBH mass distribution.

Finally, the mass function obtained with a double-peak power spectrum is bimodal, as it can be seen from \cref{fig:mass_func_double}.
Thus, the mathematical transformation $\mathcal{P}_\zeta(k)\to f(M)$ defined by the excursion-set framework is such that, if well-resolved features appear in $\mathcal{P}_\zeta(k)$, they manifest themselves in $f(M)$ as well. 
This is consistent with the physical intuition that relates comoving scales with PBH formation time and mass, but that conclusion is technically not so obvious given that the relationship between $\mathcal{P}_\zeta$ and $f(M)$ involves solving stochastic processes. 
When the two peaks have the same power-spectrum amplitude (C-2), they give rise to similar maxima in the mass distribution, while if their amplitudes are set such that they give the same contribution to $S_{\mathrm{max}}$ (C-1), the higher-mass maximum is more pronounced. 

\subsection{Cloud-in-cloud}

Let us now estimate the importance of cloud-in-cloud, namely the fact that small-mass PBHs may end up in regions forming larger-mass black holes, and therefore disappear from the final PBH census. 

\subsubsection{Comparison to Press-Schechter}

In \cref{fig:mass_func_fig} we display with dashed lines the mass function resulting from the Press-Schechter approach~\cite{Press:1973iz}, which neglects the possibility of barrier multiple crossings and thus does not account for cloud-in-cloud.
In practice, as explained in Ref.~\cite{Auclair:2020csm} and made explicit in \cref{eq:ps}, this consists of setting the triangular matrix $J$ and the kernel $K(S)$ to $0$ in \cref{eq:Volterra} and accounting for a global factor of one half.\footnote{A comparison between the excursion-set and Press-Schechter approaches at the level of the first-passage time distribution is displayed in \cref{fig:convergence-test} in \cref{sec:converge}.}
Let us emphasize that in this Press-Schechter approach, we keep the ``time'' dependence $\deltac(S)$ of the threshold in \cref{eq:x} -- whereas certain works often 
assume implicitly that the barrier is scale independent,
 $\deltac '(S) = 0$.
Physically, the Press-Schechter approach reduces to computing (one half of) the ``flux'' of trajectories across the barrier, if these trajectories are free (i.e.\ they do not feel the barrier).
As a consequence, it can very well become negative when the barrier is too steep, and such a behavior is showcased in \cref{fig:figure_6_fail}. 
This signals a clear limitation of the Press-Schechter approach when applied to systems with a moving barrier, which the excursion-set formalism overcomes.
This also explains the origin of the striking disagreement between the two approaches in the low-mass tail for models B-N1 to B-N3 and B-W1 to B-W3 in \cref{fig:mass_func_fig:bn}. 
Let us however note that the severity of the discrepancy depends on the detailed shape of the power spectrum, since the narrow top-hat spectrum is free from such pathology as displayed in \cref{fig:mass_func_fig:an}. 

The aforementioned discrepancy is not, \emph{stricto sensu}, a measure of cloud-in-cloud, but rather reflects the failure of the Press-Schechter approach in itself when the barrier $\deltac$ varies, as illustrated in \cref{fig:kernel}. 
As soon as $2S \deltac'(S) > \deltac(S)$, \cref{eq:ps} gives a nonphysical, negative value, signaling that the free flux across the barrier cannot be used as a proxy for the first-passage-time distribution.

In contrast, the excursion-set always predicts positive mass functions, with a low-mass tail that eventually becomes lighter as the amplitude of the power spectrum increases, see for instance \cref{fig:mass_func_fig:aw,fig:mass_func_fig:bw}. This is because, as the power spectrum increases, so does the number of massive PBHs, which engulf smaller ones. Light PBHs are thus less frequent since they are absorbed by their more massive companions, which is precisely what is expected from cloud-in-cloud. In the Press-Schechter approach, this effect is absent and the low-mass tails are mostly insensitive to the overall abundance.

This shows that cloud-in-cloud can play a substantial and subtle role in shaping the mass function.

\subsubsection{The role of scale separation}

\begin{figure*}[ht!]
    \centering
    \subfloat[Mass function for double log-normal models.]{\includegraphics[width=.49\textwidth]{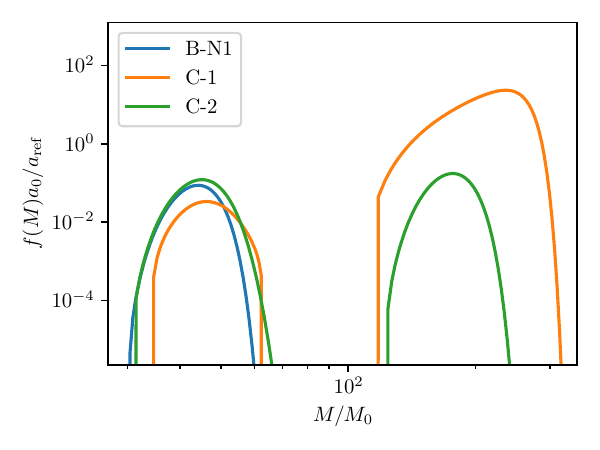} \label{fig:mass_func_double}}
    \subfloat[Probability~\eqref{eq:p2} to cross the barrier during the second stage in the piecewise model~\eqref{eq:piecewise:def}, with parameters to fit the peak of model B-N1, using $\delta_2 = 0.01$ and $S_{2} = 2 \times 10^{-5}$.
    We vary the height ($\delta_1$) and width ($S_{1}$) of the ``initial step''. 
    \label{fig:piecewise}]{\includegraphics[width=.49\textwidth]{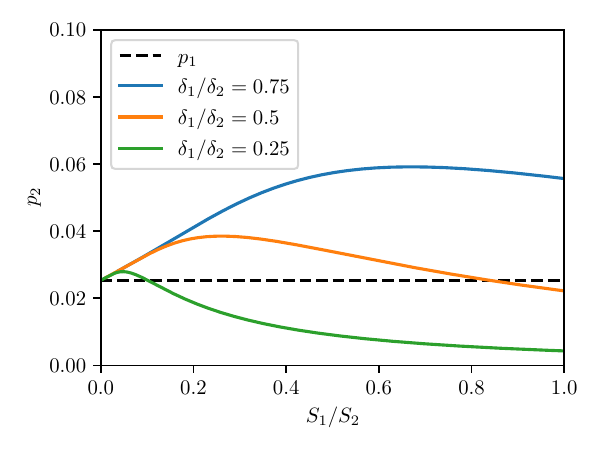}}
    \caption{Effect of a second peak at large scales on the PBH mass distribution.}
    \label{fig:figure_6c}
\end{figure*}

So far, we have focused on either quasimonochromatic or wide mass distributions, but one may wonder about the effect of cloud-in-cloud in the presence of more significant mass hierarchies. In order to discuss how heavy PBHs affect much lighter ones, in \cref{fig:mass_func_double} we start from the reference situation where a single peak at low mass is present (B-N1), then introduce a second peak at large mass (C-2), and finally increase its amplitude (C-1). One can see that, when the second large-mass peak is introduced, the abundance of low-mass PBHs slightly increases, before decreasing when the amplitude of the large-scale peak is enhanced. 

This nonmonotonic behavior can be understood as follows. If the two log-normal peaks in the power spectrum are well separated, their respective contributions to $S_{\mathrm{max}}$ read as
\begin{equation}
    \label{eq:Si}
    S_{i} = \frac{8\mathcal{A}_i}{81} \left( \frac{k_{\mathrm{p},i}}{a_0 H_0} \right)^4 e^{8\Delta^2}
\end{equation}
where $i=1,2$, see \cref{eq:Pw_lognorm}.
In that same limit, since the two peaks are narrow, $\deltac(S)$ becomes piecewise constant, with 
\begin{equation}
\label{eq:piecewise:def}
    \deltac(S) =
    \begin{cases}
        \delta_1\ \text{if}\ 0<S<S_{1} \\
        \delta_2\ \text{if}\ S_{1} <S<S_{\mathrm{max}}=S_1+S_2
    \end{cases}
\end{equation}
where we have introduced
\begin{equation}
    \label{eq:deltai}
    \delta_i = \deltac \left(\frac{ k_{\mathrm{p},i}}{a_0 H_0}\right)^2\, .
\end{equation}
In this simple piecewise toy model, as one increases the amplitude of the large-mass peak, i.e.\ $\mathcal{A}_1$,  $S_{1}$ gets larger, while the other parameters of the $\deltac(S)$ function remain fixed. This behavior is sketched across the three panels of \cref{fig:effect-of-second-peak}.

In the absence of a barrier, the transition probability for the density contrast is given by the Gaussian distribution~\eqref{eq:free}. 
The probability distribution associated to $\delta_R$ at the end of the first stage $S<S_1$ when the barrier is constant can thus be obtained by the method of images as
\begin{align}
    P(\delta_R, S_1) =& P_{\mathrm{free}}(\delta_R,S_1\vert 0) - P_{\mathrm{free}}(\delta_R,S_1\vert 2\delta_1) \nonumber \\
    = & \frac{1}{\sqrt{2\pi S_1}}\left[e^{-\frac{\delta_R^2}{2S_1}}-e^{-\frac{\left(\delta_R-2\delta_1\right)^2}{2S_1}} \right] .    
    \label{eq:P:mirror}
\end{align}
The probability to cross the barrier during the first stage is thus
\begin{equation}
    \label{eq:p1}
    p_1
    =1-\int_{-\infty}^{\delta_1}P(\delta_R,S_1)\dd{\delta_R}
    = \erfc\left(\frac{\delta_1}{\sqrt{2S_1}}\right) .
\end{equation}
A similar calculation applies to the probability to cross the barrier during the second stage, except that the initial condition for that second stage now follows the distribution~\eqref{eq:P:mirror}. This leads to  
\begin{equation}
    \label{eq:p2}
    p_2 = \int_{-\infty}^{\delta_1} P(\delta_R,S_1)\erfc\left(\frac{\delta_2-\delta_R}{\sqrt{2S_2}}\right)\dd{\delta_R}\, .
\end{equation}
The integral can be expressed in terms of the Owen's T function~\cite{Owen1980} but the corresponding expression does not bring additional insight. Keeping \cref{eq:p2} in its integral form, one can nonetheless notice that $p_2$ is nonmonotonic and exhibits two regimes.

Indeed, when $S_1\ll \delta_1^2$, the second term in \cref{eq:P:mirror} remains negligible and $p_2$ increases with $S_1$. This is because, as $S_{1}$ increases, more time becomes available for the process to climb up to the second barrier overall. In other words, at ``time'' $S_{1}$, $\delta_R$ is already displaced from the origin (it lies in the blue-shaded region in \cref{fig:effect-of-second-peak}), which makes it more likely to cross the barrier afterward. This effect is dominant for small secondary peaks, and explains why C-2 features more small-mass PBHs than B-N1.

When $S_1\gg \delta_1^2$, $P(\delta_R,S_1)$ is almost flat but its overall amplitude decays as $S_1^{-3/2}$, and so does $p_2$. In other words, the barrier may be crossed along its first (large-mass) plateau, which prevents first crossing from occurring in the second (small-mass) plateau. This is precisely the cloud-in-cloud mechanism, which dominates for larger secondary peaks, and explains why C-1 features less small-mass PBHs than C-2.

These two regimes can be observed in \cref{fig:piecewise} where $p_2$ in the piecewise model is displayed, using parameters to mimic the narrow peak of B-N1 and the situation of \cref{fig:figure_6c}. The ratio $\delta_1/\delta_2$ is a direct measure of the hierarchy of scales between the two peaks, and tends to $0$ for infinite separation. From \cref{fig:kernel-c}, one can estimate that models C-1 and C-2 have a ratio of $\delta_1/\delta_2$ between a quarter and a half. This piecewise model thus explains well why the mass function of C-1 is suppressed ($S_1 / S_2 = 1$) and the mass function of C-2 is slightly increased ($S_1 / S_2 = 1 / 16$).

\subsection{Comparison with previous estimates}

We have demonstrated in the previous section that the cloud-in-cloud mechanism may play an important role, even when different masses are involved in the mass function, as illustrated in the case of the double log-normal power spectrum.

Our results come at odds with the findings of Refs.~\cite{MoradinezhadDizgah:2019wjf, DeLuca:2020ioi}, where cloud-in-cloud was deemed to be systematically irrelevant. In these works, the argument given is that \emph{if} the formations of black holes at different masses are independent events, then the conditional probability to form a black hole of mass $M_1$ given that a black hole of mass $M_2<M_1$ already exists is $\beta(M_1\vert M_2)=\beta(M_1)$, hence it is suppressed by the PBH abundance itself. PBHs being a rare event, cloud-in-cloud should be equally rare.

Let us examine this statement more closely. In the limit where the mass ratio is large, we first note that statistical independence can indeed be recovered, provided this limit is taken while maintaining the back-hole abundances fixed. This is because, in the piecewise model described above, the abundance of the large-mass black holes is controlled by $p_1$ given in \cref{eq:p1}. As a consequence, in order to increase $M_1$ while keeping $p_1$ fixed, one needs to decrease $k_{\mathrm{p},1}$ while keeping $\delta_1/\sqrt{S_1}$ fixed. Since \cref{eq:Si,eq:deltai} indicate that $\delta_1/\sqrt{S_1}$ is independent of $k_{\mathrm{p}}$, this can be done by decreasing $k_{\mathrm{p},1}$ while keeping $\mathcal{A}_1$ fixed. When doing so, both $\delta_1$ and $S_1$ decrease, hence \cref{eq:P:mirror} asymptotes
\begin{equation}
    P\left(\delta_R,S_1\right) \underset{M_1/M_2\gg 1}{\longrightarrow} \left(1-p_1\right) \delta_{\mathrm{D}}\left(\delta_R\right) .   
\end{equation}
When substituting this expression into \cref{eq:p2}, one obtains $p_2=(1-p_1) \tilde{p}_2$, where $\tilde{p}_2=\erfc(\delta_2/\sqrt{2 S_2})$ is the abundance of light black holes that would be obtained in the absence of the large-mass population. This shows that the formation events become independent, and that cloud-in-cloud is indeed $\beta$-suppressed in that limit.

Nonetheless, a broad power spectrum is not made of well-separated, discrete peaks, and in the continuous limit, our findings show that cloud-in-cloud does occur, the wide top-hat spectrum displayed in \cref{fig:mass_func_fig:aw} being perhaps the most compelling example. It comes from the fact that, from the excursion-set perspective, the formations of PBHs at masses $M_1$ and $M_2$, i.e.\ repeated barrier crossings, are not independent events.

We have already pointed out that, as the amplitude of scalar perturbations is decreased, cloud-in-cloud may become less pronounced, especially in the case of wide power spectra.
This is because, when decreasing $\mathcal{P}_\zeta$, from \cref{eq:S'(R)}, $\vert \dd S/\dd R\vert $ decays, hence $\delta_{\mathrm{c}}'(S)\propto \vert R'(S)\vert$ increases [see e.g.\ \cref{eq:deltac(S):tophat} in the case of a top-hat power spectrum] and the barrier becomes steeper. If it grows more rapidly than the rate at which diffusion widens the $P(\delta_R)$ distribution, multiple crossings become less likely.  Therefore, at low black-hole abundances, cloud-in-cloud may be less relevant. However, the failure of the Press-Schechter approach can as well remain significant, both in the tails and in the overall amplitude of the PBH mass functions, since as previously outlined it can lead to negative abundances even for simple power spectra and irrespective of their amplitude. Moreover, since $\delta_{\mathrm{c}}'(S)$ increases in the low-amplitude limit, the constant-barrier approximation that often comes with Press-Schechter estimates may become even less valid. 
In contrast, the excursion-set method showcased in the present article 
remains applicable
for any types of shapes of power spectra, and properly accounts for correlations between PBH formation events.

\section{Conclusion}
\label{sec:Conclusion}

In this work, we have tackled some crucial yet often overlooked steps necessary to apply consistently the excursion-set formalism to 
primordial black hole formation drawing from ideas discussed in the context of metric preheating~\cite{Auclair:2020csm}.

First, we have answered the concern raised in Ref.~\cite{Kushwaha:2025zpz} pertaining to the color of the noise in the Langevin equation which describes how the coarse-grained density contrast changes when the coarse-graining scale is varied.
We have proven that, as long as Fourier modes are uncoupled, the noise is always white. However, if the excursion-set is sampled along the Hubble-crossing surface, the coarse-grained density contrast is evaluated at different times, which introduces a drift in the Langevin equation that was mistaken for a colored noise. 
Moreover, such sampling may lead to a nonmonotonic relationship between scale and density-contrast dispersion, which prevents using the latter to relabel time in the Langevin equation.
The only sampling surface that always prevents this problem from occurring is one that is synchronous.
This, however, requires to properly take into account the dependence of the barrier -- describing the PBH formation threshold -- with respect to the coarse-graining scale. 
We have demonstrated how the resulting first-passage time problem with moving barrier can be efficiently solved numerically. 
The motion of the barrier is often neglected~\cite{Dizon:2025siw, Kameli:2025qzp} but we have found that it plays a crucial role in most physical results.

Second, we have precisely reassessed under which conditions the cloud-in-cloud phenomenon becomes relevant.
While it is innocuous in the limit of a large separation of scales (i.e.\ for black holes forming at two widely different masses), it is more substantial in mass functions arising for broad power spectra where a continuous set of scales is enhanced. It also exhibits a subtle dependence on the relative ratio of powers: at first, increasing the power at large scales (i.e.\ large mass) leads to a moderate enhancement of small-mass black holes due to the spread of the initial conditions in the second stage of the excursion set; then, as the power at large scale continues to increase, the low-mass end of the mass fraction inevitably decays.
This suggests that cloud-in-cloud is not always as irrelevant as argued in Refs.~\cite{MoradinezhadDizgah:2019wjf, DeLuca:2020ioi}.

Third, even when cloud-in-cloud does not play a significant role, we have put forward the relevance of the excursion-set approach compared 
to the Press-Schechter approach. The latter,  with a scale-dependent threshold -- setting $\deltac'(S)=0$ only worsens the results -- fails, for most models, to predict both the amplitude and the low-mass tails of the PBH mass function. In certain cases, it even predicts negative abundances.

Let us finally stress that the excursion set rests on a few assumptions beyond which it remains to be extended. The main one is probably that the density field obeys Gaussian statistics, while large fluctuations are known to display strong non-Gaussian features with heavy tails~\cite{Pattison:2017mbe, Franciolini:2018vbk, Ezquiaga:2019ftu, Ando:2020fjm, Figueroa:2020jkf, Tada:2021zzj, Kitajima:2021fpq, Hooshangi:2021ubn, Biagetti:2021eep, Gow:2022jfb, Raatikainen:2023bzk, Firouzjahi:2023xke, Animali:2024jiz, Vennin:2024yzl, Jackson:2024aoo, Ianniccari:2024bkh, Animali:2025pyf, 2025zndo..15235932A, Choudhury:2025kxg}. Other aspects of the PBH formation should be included in the formalism, such as the role of critical scaling~\cite{Niemeyer:1997mt}, or the use of the compaction function~\cite{Shibata:1999zs} and refined window functions (as recently discussed in Ref.~\cite{Saito:2025sny}) to improve the reliability of the collapse criterion.

\begin{acknowledgments}
P.A. and V.V. acknowledge the SNCF REMI train service between Paris and Orl\'eans for hosting multiple (often longer than scheduled) discussions while this work was in progress.
B.B. is publishing in the quality of ASPIRANT Research Fellow of the ``Fonds de la Recherche Scientifique - FNRS''.
We also thank Ashu Kushwaha and Teruaki Suyama for interesting discussions on their work.

\end{acknowledgments}

\appendix

\newpage
\onecolumngrid

\section{Numerical tips}
\label{sec:num_tips}

As explained in \cref{sec:FPT}, we find the first-passage time distribution $P_\mathrm{FPT}$ by solving the linear matrix equation \eqref{eq:Volterra}.
From this distribution, we obtain the transition probability by performing the integral described in \cref{eq:proba_transition},
\begin{equation}
    P(\delta_R, S \vert \delta_{R,\mathrm{i}}, S_{\mathrm{i}} ) = P_{\mathrm{free}}(\delta_R, S | \delta_{R,\mathrm{i}}, S_{\mathrm{i}}) - \int_{0}^{S} \dd{s} P_{\mathrm{FPT}}(s | \delta_{R,\mathrm{i}}, S_{\mathrm{i}}) P_{\mathrm{free}} \left[ \delta_R - \deltac(s), S-s | \delta_{R,\mathrm{i}}, S_{\mathrm{i}}   \right] .
\end{equation}
By construction, this probability vanishes on the boundary, i.e.\ $P[\deltac(S), S \vert \delta_{R,\mathrm{i}}, S_{\mathrm{i}}] = 0$ and 
\begin{equation}
    0 = P_{\mathrm{free}}(\deltac(S), S | \delta_{R,\mathrm{i}}, S_{\mathrm{i}}) - \int_{0}^{S} \dd{s} P_{\mathrm{FPT}}(s | \delta_{R,\mathrm{i}}, S_{\mathrm{i}}) P_{\mathrm{free}} \left[ \deltac(S) - \deltac(s), S-s | \delta_{R,\mathrm{i}}, S_{\mathrm{i}}   \right].
    \label{eq:proba_transition_boundary}
\end{equation}
As a consequence, solving \cref{eq:proba_transition} close to the boundary $\delta_R \to \deltac(S)$  yields a vanishing result and is subject to truncation errors.
Nonetheless, one can combine \cref{eq:proba_transition} with \cref{eq:proba_transition_boundary} to obtain an equation that is immune to this type of truncation error
\begin{multline}
    P(\delta_R, S \vert \delta_{R,\mathrm{i}}, S_{\mathrm{i}} ) 
    = \int_{0}^{S} \dd{s} P_{\mathrm{FPT}}(s | \delta_{R,\mathrm{i}}, S_{\mathrm{i}}) \\
    \cross \left\{ 
    \frac{P_{\mathrm{free}}(\delta_R, S | \delta_{R,\mathrm{i}}, S_{\mathrm{i}})}{P_{\mathrm{free}}(\deltac(S), S | \delta_{R,\mathrm{i}}, S_{\mathrm{i}})}
    P_{\mathrm{free}} \left[ \deltac(S) - \deltac(s), S-s | \delta_{R,\mathrm{i}}, S_{\mathrm{i}}   \right]
    - P_{\mathrm{free}} \left[ \delta_R - \deltac(s), S-s | \delta_{R,\mathrm{i}}, S_{\mathrm{i}} \right] 
    \right\} .
\end{multline}

We also report that our results for the first-passage-time distribution converge faster if the discretization is performed on the logarithm of the variance $S = S_\mathrm{min} \exp(n \Delta \ln s)$ instead of the variance directly. 
In this case, the scalar $\Delta s$ in \cref{eq:matrix-eq} is promoted to a diagonal matrix that multiplies $M$. 

\section{Consistency checks}
\label{sec:Discussion}

\begin{figure*}[t]
    \centering
    \subfloat[Impact of the reference time on the mass function.]{\includegraphics[width=0.49\textwidth]{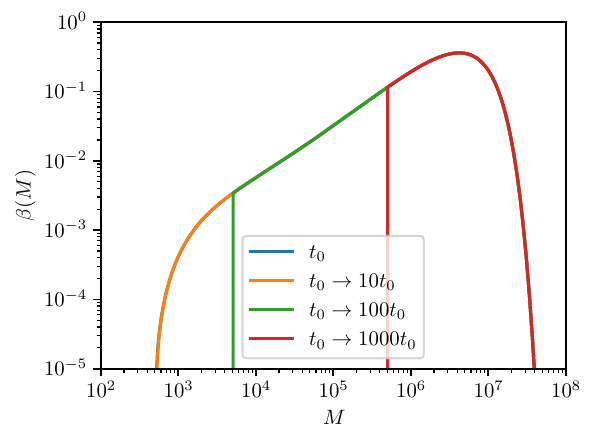}\label{fig:uv}}
    \subfloat[Impact of the IR cutoff on the mass function]{\includegraphics[width=0.49\textwidth]{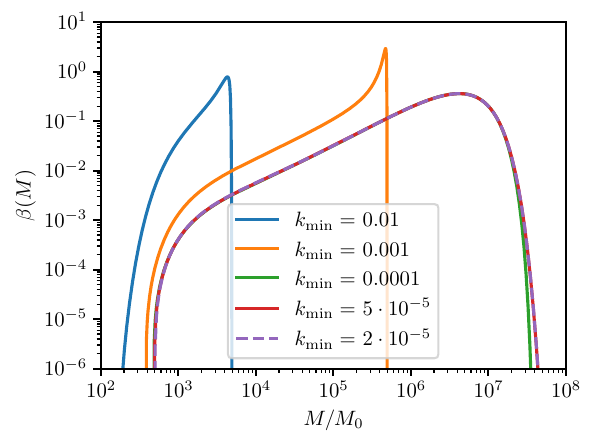}\label{fig:ir}}
    \caption{Impact of the reference time and of the IR cutoff $k_\mathrm{min}$ on the mass function $\beta(M)$ for the benchmark model B-W1. Masses are expressed in units of the initial $M_0$. On the left panel, all the curves superimpose exactly.
    }
    \label{fig:ir_uv}
\end{figure*}

\begin{figure*}[t]
    \centering
    \subfloat[$\delta_{R,\mathrm{i}} = 0, S_{\mathrm{i}} = 0$]{\includegraphics[width=0.49\textwidth]{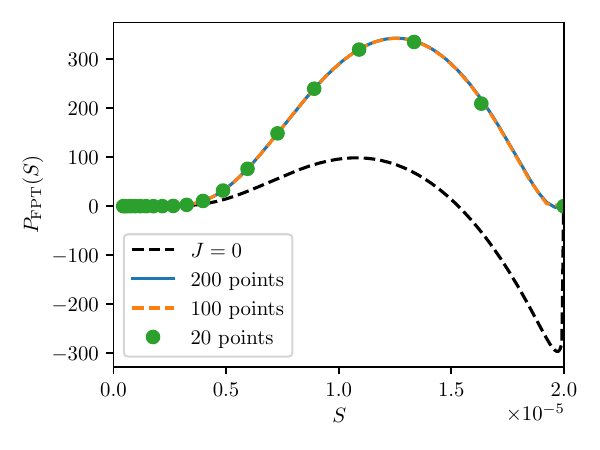}} %
    \subfloat[$\delta_{R,\mathrm{i}} = -0.005, S_{\mathrm{i}} = 0$]{\includegraphics[width=0.49\textwidth]{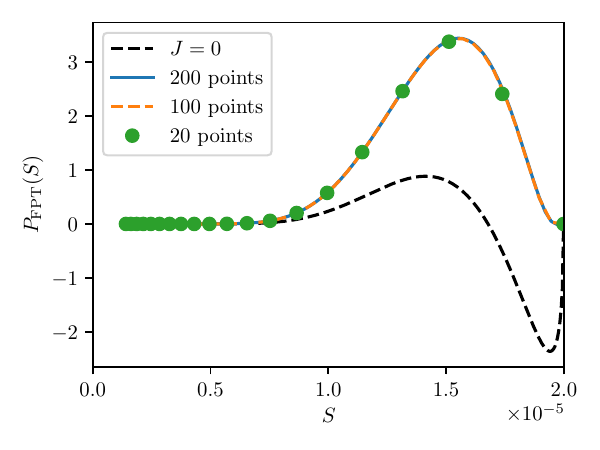}}\\
    \subfloat[$\delta_{R,\mathrm{i}} = 0, S_{\mathrm{i}} = 10^{-5}$]{\includegraphics[width=0.49\textwidth]{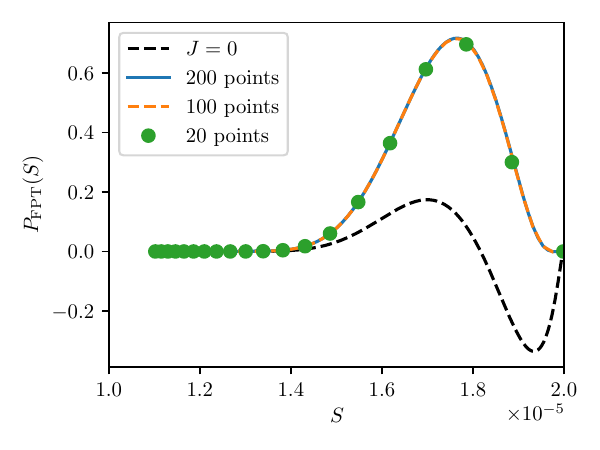}} %
    \subfloat[$\delta_{R,\mathrm{i}} = -0.005, S_{\mathrm{i}} = 10^{-5}$]{\includegraphics[width=0.49\textwidth]{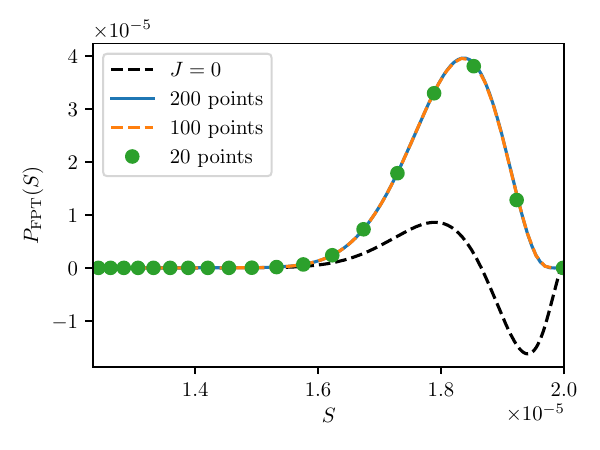}} %
    \caption{
        Convergence test for model B-N1 and an increasing number of points in the discretization of \cref{eq:Volterra}. We selected different starting points $(\delta_{R,\mathrm{i}}, S_{\mathrm{i}})$ for each panel.
    }
    \label{fig:convergence-test}
\end{figure*}

\subsection{Dependence on the reference time}
\label{sec:UV}

As explained in \cref{subsec:synchronous}, 
when the excursion-set is performed on a synchronous hypersurface, a certain reference time $t_0$ must be chosen, which amounts to setting a certain wave number $k_\mathrm{max}=a_0 H_0$.  A natural choice for $t_0$ can be the end of inflation~\cite{Auclair:2020csm}, such that all scales amplified by inflation are included in the analysis, although
any time preceding the reentry of the peak of the curvature perturbations power spectrum is suitable. Indeed, if scales reentering before the peak are suppressed, they have little impact on PBH formation, hence physical results do not depend on $t_0$ provided it is taken early enough.
To prove this, let us change the reference time $t_0\to t_1$.
From \cref{eq:deltac:rescaling}, this leads to a rescaling of the barrier by a constant multiplicative factor,
\begin{equation}
    \deltac(S)\to A \deltac(S) \quad \text{where}\quad A=\left(\frac{H_1}{H_0}\right)^{-\frac{2}{3}\frac{1+3w}{1+w}}\, .
\end{equation}
Meanwhile, since $S(R,t)$ involves $P_\delta(k,t)$, see \cref{eq:S_and_R}, and given that $\delta_{\vb{k}}(t) \propto (aH)^{-2}\propto H^{-\frac{2}{3}\frac{1+3w}{1+w}}$ at super-Hubble scales, one also needs to rescale
\begin{equation}
    S_{\mathrm{max}}\to A^2 S_{\mathrm{max}}\, .
\end{equation}
It is clear that the stochastic process~\eqref{eq:langevin:synchronous:S} is invariant under the rescaling $\delta_R\to A\delta_R$ and $S\to A^2 S$, since this amounts to rescaling both $\dd \delta_R/\dd S$ and $\xi(S)$ by the same factor $A^{-1}$. As a consequence, the mass function is not affected by the choice of $t_0$, at least for masses above $M_0$.
This is confirmed in \cref{fig:uv} in the case of a log-normal power spectrum where we show that different choices of $t_0$ do not affect the mass function but only change the range of scales that are selected. If $t_0$ is chosen after the reentry of the peak scale of the power spectrum, 
the mass function is not fully reconstructed.

\subsection{Dependence on the IR cutoff}
\label{subsec:IR}

As mentioned in \cref{subsec:lognorm}, in the case of the log-normal power spectrum,
an infrared cutoff $k_\mathrm{min}$ is needed for all the quantities to be numerically well behaved.
For the top-hat power spectrum of \cref{subsec:broad}, such a scale arises by construction. As for the reference time discussed above -- and that may be seen as an ultraviolet cutoff -- this IR cutoff does not impact our results, since the density contrast is $k^2$ suppressed with respect to the curvature perturbation. 
Consequently, adding up or discarding IR modes (by decreasing or increasing $k_\mathrm{min}$) does not impact much the integral of \cref{eq:S(R)} defining $S(R)$. 
As such, $k_\mathrm{min}$ only plays the role of 
a numerical regularization scheme, that should be evaluated to sufficiently low values such that final results do not depend on it.
(in practice it has been fixed to $10^{-3}$ or $10^{-5}$ in our applications, see \cref{tab:lognorm}).
This is confirmed in \cref{fig:ir}, in the case of a wide log-normal power spectrum. 
As soon as $k_\mathrm{min}$ is of order $10^{-4}$ or smaller, that is only one order of magnitude smaller than the wave number $k_\mathrm{p}$ at which the power spectrum peaks, the mass function converges to the same result.

\section{Convergence test for the log-normal model}
\label{sec:converge}

In \cref{fig:convergence-test}, we perform a convergence test for the first-passage distribution by increasing the number of points in the discretization of the Volterra integral equation \eqref{eq:Volterra} for model B-W1. 
We find overall that the convergence of our code is reached very rapidly and consistently for all models.
Above $100$ points, the results are basically indistinguishable from one another.
Surprisingly, we recover accurate results with as little as $20$ points in the discretization, although the resulting distribution would be too sparse to be used in practice.
In this article, all the results are presented with $1000$ points in the discretization.

\bibstyle{aps}
\bibliography{biblio}

\begin{thebibliography}{96}%
\makeatletter
\providecommand \@ifxundefined [1]{%
 \@ifx{#1\undefined}
}%
\providecommand \@ifnum [1]{%
 \ifnum #1\expandafter \@firstoftwo
 \else \expandafter \@secondoftwo
 \fi
}%
\providecommand \@ifx [1]{%
 \ifx #1\expandafter \@firstoftwo
 \else \expandafter \@secondoftwo
 \fi
}%
\providecommand \natexlab [1]{#1}%
\providecommand \enquote  [1]{``#1''}%
\providecommand \bibnamefont  [1]{#1}%
\providecommand \bibfnamefont [1]{#1}%
\providecommand \citenamefont [1]{#1}%
\providecommand \href@noop [0]{\@secondoftwo}%
\providecommand \href [0]{\begingroup \@sanitize@url \@href}%
\providecommand \@href[1]{\@@startlink{#1}\@@href}%
\providecommand \@@href[1]{\endgroup#1\@@endlink}%
\providecommand \@sanitize@url [0]{\catcode `\\12\catcode `\$12\catcode
  `\&12\catcode `\#12\catcode `\^12\catcode `\_12\catcode `\%12\relax}%
\providecommand \@@startlink[1]{}%
\providecommand \@@endlink[0]{}%
\providecommand \url  [0]{\begingroup\@sanitize@url \@url }%
\providecommand \@url [1]{\endgroup\@href {#1}{\urlprefix }}%
\providecommand \urlprefix  [0]{URL }%
\providecommand \Eprint [0]{\href }%
\providecommand \doibase [0]{http://dx.doi.org/}%
\providecommand \selectlanguage [0]{\@gobble}%
\providecommand \bibinfo  [0]{\@secondoftwo}%
\providecommand \bibfield  [0]{\@secondoftwo}%
\providecommand \translation [1]{[#1]}%
\providecommand \BibitemOpen [0]{}%
\providecommand \bibitemStop [0]{}%
\providecommand \bibitemNoStop [0]{.\EOS\space}%
\providecommand \EOS [0]{\spacefactor3000\relax}%
\providecommand \BibitemShut  [1]{\csname bibitem#1\endcsname}%
\let\auto@bib@innerbib\@empty
\bibitem [{\citenamefont {Carr}\ and\ \citenamefont
  {Hawking}(1974)}]{Carr:1974nx}%
  \BibitemOpen
  \bibfield  {author} {\bibinfo {author} {\bibfnamefont {B.~J.}\ \bibnamefont
  {Carr}}\ and\ \bibinfo {author} {\bibfnamefont {S.~W.}\ \bibnamefont
  {Hawking}},\ }\href {\doibase 10.1093/mnras/168.2.399} {\bibfield  {journal}
  {\bibinfo  {journal} {Mon. Not. Roy. Astron. Soc.}\ }\textbf {\bibinfo
  {volume} {168}},\ \bibinfo {pages} {399} (\bibinfo {year}
  {1974})}\BibitemShut {NoStop}%
\bibitem [{\citenamefont {Carr}(1975)}]{Carr:1975qj}%
  \BibitemOpen
  \bibfield  {author} {\bibinfo {author} {\bibfnamefont {B.~J.}\ \bibnamefont
  {Carr}},\ }\href {\doibase 10.1086/153853} {\bibfield  {journal} {\bibinfo
  {journal} {Astrophys. J.}\ }\textbf {\bibinfo {volume} {201}},\ \bibinfo
  {pages} {1} (\bibinfo {year} {1975})}\BibitemShut {NoStop}%
\bibitem [{\citenamefont {Chapline}(1975)}]{Chapline:1975ojl}%
  \BibitemOpen
  \bibfield  {author} {\bibinfo {author} {\bibfnamefont {G.~F.}\ \bibnamefont
  {Chapline}},\ }\href {\doibase 10.1038/253251a0} {\bibfield  {journal}
  {\bibinfo  {journal} {Nature}\ }\textbf {\bibinfo {volume} {253}},\ \bibinfo
  {pages} {251} (\bibinfo {year} {1975})}\BibitemShut {NoStop}%
\bibitem [{\citenamefont {Meszaros}(1975)}]{Meszaros:1975ef}%
  \BibitemOpen
  \bibfield  {author} {\bibinfo {author} {\bibfnamefont {P.}~\bibnamefont
  {Meszaros}},\ }\href@noop {} {\bibfield  {journal} {\bibinfo  {journal}
  {Astron. Astrophys.}\ }\textbf {\bibinfo {volume} {38}},\ \bibinfo {pages}
  {5} (\bibinfo {year} {1975})}\BibitemShut {NoStop}%
\bibitem [{\citenamefont {Duechting}(2004)}]{Duechting:2004dk}%
  \BibitemOpen
  \bibfield  {author} {\bibinfo {author} {\bibfnamefont {N.}~\bibnamefont
  {Duechting}},\ }\href {\doibase 10.1103/PhysRevD.70.064015} {\bibfield
  {journal} {\bibinfo  {journal} {Phys. Rev. D}\ }\textbf {\bibinfo {volume}
  {70}},\ \bibinfo {pages} {064015} (\bibinfo {year} {2004})},\ \Eprint
  {http://arxiv.org/abs/astro-ph/0406260} {arXiv:astro-ph/0406260} \BibitemShut
  {NoStop}%
\bibitem [{\citenamefont {Kawasaki}\ \emph {et~al.}(2012)\citenamefont
  {Kawasaki}, \citenamefont {Kusenko},\ and\ \citenamefont
  {Yanagida}}]{Kawasaki:2012kn}%
  \BibitemOpen
  \bibfield  {author} {\bibinfo {author} {\bibfnamefont {M.}~\bibnamefont
  {Kawasaki}}, \bibinfo {author} {\bibfnamefont {A.}~\bibnamefont {Kusenko}}, \
  and\ \bibinfo {author} {\bibfnamefont {T.~T.}\ \bibnamefont {Yanagida}},\
  }\href {\doibase 10.1016/j.physletb.2012.03.056} {\bibfield  {journal}
  {\bibinfo  {journal} {Phys. Lett. B}\ }\textbf {\bibinfo {volume} {711}},\
  \bibinfo {pages} {1} (\bibinfo {year} {2012})},\ \Eprint
  {http://arxiv.org/abs/1202.3848} {arXiv:1202.3848 [astro-ph.CO]} \BibitemShut
  {NoStop}%
\bibitem [{\citenamefont {Chiba}\ and\ \citenamefont
  {Yokoyama}(2017)}]{Chiba:2017rvs}%
  \BibitemOpen
  \bibfield  {author} {\bibinfo {author} {\bibfnamefont {T.}~\bibnamefont
  {Chiba}}\ and\ \bibinfo {author} {\bibfnamefont {S.}~\bibnamefont
  {Yokoyama}},\ }\href {\doibase 10.1093/ptep/ptx087} {\bibfield  {journal}
  {\bibinfo  {journal} {PTEP}\ }\textbf {\bibinfo {volume} {2017}},\ \bibinfo
  {pages} {083E01} (\bibinfo {year} {2017})},\ \Eprint
  {http://arxiv.org/abs/1704.06573} {arXiv:1704.06573 [gr-qc]} \BibitemShut
  {NoStop}%
\bibitem [{\citenamefont {Bagui}\ \emph {et~al.}(2025)\citenamefont {Bagui}
  \emph {et~al.}}]{LISACosmologyWorkingGroup:2023njw}%
  \BibitemOpen
  \bibfield  {author} {\bibinfo {author} {\bibfnamefont {E.}~\bibnamefont
  {Bagui}} \emph {et~al.} (\bibinfo {collaboration} {LISA Cosmology Working
  Group}),\ }\href {\doibase 10.1007/s41114-024-00053-w} {\bibfield  {journal}
  {\bibinfo  {journal} {Living Rev. Rel.}\ }\textbf {\bibinfo {volume} {28}},\
  \bibinfo {pages} {1} (\bibinfo {year} {2025})},\ \Eprint
  {http://arxiv.org/abs/2310.19857} {arXiv:2310.19857 [astro-ph.CO]}
  \BibitemShut {NoStop}%
\bibitem [{\citenamefont {Antoniadis}\ \emph {et~al.}(2024)\citenamefont
  {Antoniadis} \emph {et~al.}}]{EPTA:2023xxk}%
  \BibitemOpen
  \bibfield  {author} {\bibinfo {author} {\bibfnamefont {J.}~\bibnamefont
  {Antoniadis}} \emph {et~al.} (\bibinfo {collaboration} {EPTA, InPTA}),\
  }\href {\doibase 10.1051/0004-6361/202347433} {\bibfield  {journal} {\bibinfo
   {journal} {Astron. Astrophys.}\ }\textbf {\bibinfo {volume} {685}},\
  \bibinfo {pages} {A94} (\bibinfo {year} {2024})},\ \Eprint
  {http://arxiv.org/abs/2306.16227} {arXiv:2306.16227 [astro-ph.CO]}
  \BibitemShut {NoStop}%
\bibitem [{\citenamefont {Afzal}\ \emph {et~al.}(2023)\citenamefont {Afzal}
  \emph {et~al.}}]{NANOGrav:2023hvm}%
  \BibitemOpen
  \bibfield  {author} {\bibinfo {author} {\bibfnamefont {A.}~\bibnamefont
  {Afzal}} \emph {et~al.} (\bibinfo {collaboration} {NANOGrav collaboration}),\
  }\href {\doibase 10.3847/2041-8213/acdc91} {\bibfield  {journal} {\bibinfo
  {journal} {The Astrophysical Journal Letters}\ }\textbf {\bibinfo {volume}
  {951}},\ \bibinfo {pages} {L11} (\bibinfo {year} {2023})},\ \Eprint
  {http://arxiv.org/abs/2306.16219} {2306.16219} \BibitemShut {NoStop}%
\bibitem [{\citenamefont {Niikura}\ \emph
  {et~al.}(2019{\natexlab{a}})\citenamefont {Niikura} \emph
  {et~al.}}]{Niikura:2017zjd}%
  \BibitemOpen
  \bibfield  {author} {\bibinfo {author} {\bibfnamefont {H.}~\bibnamefont
  {Niikura}} \emph {et~al.},\ }\href {\doibase 10.1038/s41550-019-0723-1}
  {\bibfield  {journal} {\bibinfo  {journal} {Nature Astron.}\ }\textbf
  {\bibinfo {volume} {3}},\ \bibinfo {pages} {524} (\bibinfo {year}
  {2019}{\natexlab{a}})},\ \Eprint {http://arxiv.org/abs/1701.02151}
  {arXiv:1701.02151 [astro-ph.CO]} \BibitemShut {NoStop}%
\bibitem [{\citenamefont {Niikura}\ \emph
  {et~al.}(2019{\natexlab{b}})\citenamefont {Niikura}, \citenamefont {Takada},
  \citenamefont {Yokoyama}, \citenamefont {Sumi},\ and\ \citenamefont
  {Masaki}}]{Niikura:2019kqi}%
  \BibitemOpen
  \bibfield  {author} {\bibinfo {author} {\bibfnamefont {H.}~\bibnamefont
  {Niikura}}, \bibinfo {author} {\bibfnamefont {M.}~\bibnamefont {Takada}},
  \bibinfo {author} {\bibfnamefont {S.}~\bibnamefont {Yokoyama}}, \bibinfo
  {author} {\bibfnamefont {T.}~\bibnamefont {Sumi}}, \ and\ \bibinfo {author}
  {\bibfnamefont {S.}~\bibnamefont {Masaki}},\ }\href {\doibase
  10.1103/PhysRevD.99.083503} {\bibfield  {journal} {\bibinfo  {journal} {Phys.
  Rev. D}\ }\textbf {\bibinfo {volume} {99}},\ \bibinfo {pages} {083503}
  (\bibinfo {year} {2019}{\natexlab{b}})},\ \Eprint
  {http://arxiv.org/abs/1901.07120} {arXiv:1901.07120 [astro-ph.CO]}
  \BibitemShut {NoStop}%
\bibitem [{\citenamefont {Sugiyama}\ \emph {et~al.}(2023)\citenamefont
  {Sugiyama}, \citenamefont {Takada},\ and\ \citenamefont
  {Kusenko}}]{Sugiyama:2021xqg}%
  \BibitemOpen
  \bibfield  {author} {\bibinfo {author} {\bibfnamefont {S.}~\bibnamefont
  {Sugiyama}}, \bibinfo {author} {\bibfnamefont {M.}~\bibnamefont {Takada}}, \
  and\ \bibinfo {author} {\bibfnamefont {A.}~\bibnamefont {Kusenko}},\ }\href
  {\doibase 10.1016/j.physletb.2023.137891} {\bibfield  {journal} {\bibinfo
  {journal} {Phys. Lett. B}\ }\textbf {\bibinfo {volume} {840}},\ \bibinfo
  {pages} {137891} (\bibinfo {year} {2023})},\ \Eprint
  {http://arxiv.org/abs/2108.03063} {arXiv:2108.03063 [hep-ph]} \BibitemShut
  {NoStop}%
\bibitem [{\citenamefont {Escriv{\`a}}\ \emph {et~al.}(2022)\citenamefont
  {Escriv{\`a}}, \citenamefont {Kuhnel},\ and\ \citenamefont
  {Tada}}]{Escriva:2022duf}%
  \BibitemOpen
  \bibfield  {author} {\bibinfo {author} {\bibfnamefont {A.}~\bibnamefont
  {Escriv{\`a}}}, \bibinfo {author} {\bibfnamefont {F.}~\bibnamefont {Kuhnel}},
  \ and\ \bibinfo {author} {\bibfnamefont {Y.}~\bibnamefont {Tada}},\ }\href
  {\doibase 10.1016/B978-0-32-395636-9.00012-8} {\  (\bibinfo {year} {2022}),\
  10.1016/B978-0-32-395636-9.00012-8},\ \Eprint
  {http://arxiv.org/abs/2211.05767} {arXiv:2211.05767 [astro-ph.CO]}
  \BibitemShut {NoStop}%
\bibitem [{\citenamefont {Press}\ and\ \citenamefont
  {Schechter}(1974)}]{Press:1973iz}%
  \BibitemOpen
  \bibfield  {author} {\bibinfo {author} {\bibfnamefont {W.~H.}\ \bibnamefont
  {Press}}\ and\ \bibinfo {author} {\bibfnamefont {P.}~\bibnamefont
  {Schechter}},\ }\href {\doibase 10.1086/152650} {\bibfield  {journal}
  {\bibinfo  {journal} {Astrophys. J.}\ }\textbf {\bibinfo {volume} {187}},\
  \bibinfo {pages} {425} (\bibinfo {year} {1974})}\BibitemShut {NoStop}%
\bibitem [{\citenamefont {Bardeen}\ \emph {et~al.}(1986)\citenamefont
  {Bardeen}, \citenamefont {Bond}, \citenamefont {Kaiser},\ and\ \citenamefont
  {Szalay}}]{Bardeen:1985tr}%
  \BibitemOpen
  \bibfield  {author} {\bibinfo {author} {\bibfnamefont {J.~M.}\ \bibnamefont
  {Bardeen}}, \bibinfo {author} {\bibfnamefont {J.~R.}\ \bibnamefont {Bond}},
  \bibinfo {author} {\bibfnamefont {N.}~\bibnamefont {Kaiser}}, \ and\ \bibinfo
  {author} {\bibfnamefont {A.~S.}\ \bibnamefont {Szalay}},\ }\href {\doibase
  10.1086/164143} {\bibfield  {journal} {\bibinfo  {journal} {Astrophys. J.}\
  }\textbf {\bibinfo {volume} {304}},\ \bibinfo {pages} {15} (\bibinfo {year}
  {1986})}\BibitemShut {NoStop}%
\bibitem [{\citenamefont {Young}\ and\ \citenamefont
  {Musso}(2020)}]{Young:2020xmk}%
  \BibitemOpen
  \bibfield  {author} {\bibinfo {author} {\bibfnamefont {S.}~\bibnamefont
  {Young}}\ and\ \bibinfo {author} {\bibfnamefont {M.}~\bibnamefont {Musso}},\
  }\href {\doibase 10.1088/1475-7516/2020/11/022} {\bibfield  {journal}
  {\bibinfo  {journal} {JCAP}\ }\textbf {\bibinfo {volume} {11}},\ \bibinfo
  {pages} {022} (\bibinfo {year} {2020})},\ \Eprint
  {http://arxiv.org/abs/2001.06469} {arXiv:2001.06469 [astro-ph.CO]}
  \BibitemShut {NoStop}%
\bibitem [{\citenamefont {Inman}\ and\ \citenamefont
  {Ali-Ha{\"\i}moud}(2019)}]{Inman:2019wvr}%
  \BibitemOpen
  \bibfield  {author} {\bibinfo {author} {\bibfnamefont {D.}~\bibnamefont
  {Inman}}\ and\ \bibinfo {author} {\bibfnamefont {Y.}~\bibnamefont
  {Ali-Ha{\"\i}moud}},\ }\href {\doibase 10.1103/PhysRevD.100.083528}
  {\bibfield  {journal} {\bibinfo  {journal} {Phys. Rev. D}\ }\textbf {\bibinfo
  {volume} {100}},\ \bibinfo {pages} {083528} (\bibinfo {year} {2019})},\
  \Eprint {http://arxiv.org/abs/1907.08129} {arXiv:1907.08129 [astro-ph.CO]}
  \BibitemShut {NoStop}%
\bibitem [{\citenamefont {Tkachev}\ \emph {et~al.}(2020)\citenamefont
  {Tkachev}, \citenamefont {Pilipenko},\ and\ \citenamefont
  {Yepes}}]{Tkachev:2020uin}%
  \BibitemOpen
  \bibfield  {author} {\bibinfo {author} {\bibfnamefont {M.}~\bibnamefont
  {Tkachev}}, \bibinfo {author} {\bibfnamefont {S.}~\bibnamefont {Pilipenko}},
  \ and\ \bibinfo {author} {\bibfnamefont {G.}~\bibnamefont {Yepes}},\ }\href
  {\doibase 10.1093/mnras/staa3103} {\bibfield  {journal} {\bibinfo  {journal}
  {Mon. Not. Roy. Astron. Soc.}\ }\textbf {\bibinfo {volume} {499}},\ \bibinfo
  {pages} {4854} (\bibinfo {year} {2020})},\ \Eprint
  {http://arxiv.org/abs/2009.07813} {arXiv:2009.07813 [astro-ph.CO]}
  \BibitemShut {NoStop}%
\bibitem [{\citenamefont {Trashorras}\ \emph {et~al.}(2021)\citenamefont
  {Trashorras}, \citenamefont {Garc\'\i{}a-Bellido},\ and\ \citenamefont
  {Nesseris}}]{Trashorras:2020mwn}%
  \BibitemOpen
  \bibfield  {author} {\bibinfo {author} {\bibfnamefont {M.}~\bibnamefont
  {Trashorras}}, \bibinfo {author} {\bibfnamefont {J.}~\bibnamefont
  {Garc\'\i{}a-Bellido}}, \ and\ \bibinfo {author} {\bibfnamefont
  {S.}~\bibnamefont {Nesseris}},\ }\href {\doibase 10.3390/universe7010018}
  {\bibfield  {journal} {\bibinfo  {journal} {Universe}\ }\textbf {\bibinfo
  {volume} {7}},\ \bibinfo {pages} {18} (\bibinfo {year} {2021})},\ \Eprint
  {http://arxiv.org/abs/2006.15018} {arXiv:2006.15018 [astro-ph.CO]}
  \BibitemShut {NoStop}%
\bibitem [{\citenamefont {Peacock}\ and\ \citenamefont
  {Heavens}(1990)}]{Peacock:1990zz}%
  \BibitemOpen
  \bibfield  {author} {\bibinfo {author} {\bibfnamefont {J.~A.}\ \bibnamefont
  {Peacock}}\ and\ \bibinfo {author} {\bibfnamefont {A.~F.}\ \bibnamefont
  {Heavens}},\ }\href@noop {} {\bibfield  {journal} {\bibinfo  {journal} {Mon.
  Not. Roy. Astron. Soc.}\ }\textbf {\bibinfo {volume} {243}},\ \bibinfo
  {pages} {133} (\bibinfo {year} {1990})}\BibitemShut {NoStop}%
\bibitem [{\citenamefont {Bower}(1991)}]{Bower:1991kf}%
  \BibitemOpen
  \bibfield  {author} {\bibinfo {author} {\bibfnamefont {R.~G.}\ \bibnamefont
  {Bower}},\ }\href@noop {} {\bibfield  {journal} {\bibinfo  {journal} {Mon.
  Not. Roy. Astron. Soc.}\ }\textbf {\bibinfo {volume} {248}},\ \bibinfo
  {pages} {332} (\bibinfo {year} {1991})}\BibitemShut {NoStop}%
\bibitem [{\citenamefont {Bond}\ \emph {et~al.}(1991)\citenamefont {Bond},
  \citenamefont {Cole}, \citenamefont {Efstathiou},\ and\ \citenamefont
  {Kaiser}}]{Bond:1990iw}%
  \BibitemOpen
  \bibfield  {author} {\bibinfo {author} {\bibfnamefont {J.~R.}\ \bibnamefont
  {Bond}}, \bibinfo {author} {\bibfnamefont {S.}~\bibnamefont {Cole}}, \bibinfo
  {author} {\bibfnamefont {G.}~\bibnamefont {Efstathiou}}, \ and\ \bibinfo
  {author} {\bibfnamefont {N.}~\bibnamefont {Kaiser}},\ }\href {\doibase
  10.1086/170520} {\bibfield  {journal} {\bibinfo  {journal} {Astrophys. J.}\
  }\textbf {\bibinfo {volume} {379}},\ \bibinfo {pages} {440} (\bibinfo {year}
  {1991})}\BibitemShut {NoStop}%
\bibitem [{\citenamefont {Shibata}\ and\ \citenamefont
  {Sasaki}(1999)}]{Shibata:1999zs}%
  \BibitemOpen
  \bibfield  {author} {\bibinfo {author} {\bibfnamefont {M.}~\bibnamefont
  {Shibata}}\ and\ \bibinfo {author} {\bibfnamefont {M.}~\bibnamefont
  {Sasaki}},\ }\href {\doibase 10.1103/PhysRevD.60.084002} {\bibfield
  {journal} {\bibinfo  {journal} {Phys. Rev. D}\ }\textbf {\bibinfo {volume}
  {60}},\ \bibinfo {pages} {084002} (\bibinfo {year} {1999})},\ \Eprint
  {http://arxiv.org/abs/gr-qc/9905064} {arXiv:gr-qc/9905064} \BibitemShut
  {NoStop}%
\bibitem [{\citenamefont {Musco}(2019)}]{Musco:2018rwt}%
  \BibitemOpen
  \bibfield  {author} {\bibinfo {author} {\bibfnamefont {I.}~\bibnamefont
  {Musco}},\ }\href {\doibase 10.1103/PhysRevD.100.123524} {\bibfield
  {journal} {\bibinfo  {journal} {Phys. Rev. D}\ }\textbf {\bibinfo {volume}
  {100}},\ \bibinfo {pages} {123524} (\bibinfo {year} {2019})},\ \Eprint
  {http://arxiv.org/abs/1809.02127} {arXiv:1809.02127 [gr-qc]} \BibitemShut
  {NoStop}%
\bibitem [{\citenamefont {Escriv\`a}\ \emph {et~al.}(2020)\citenamefont
  {Escriv\`a}, \citenamefont {Germani},\ and\ \citenamefont
  {Sheth}}]{Escriva:2019phb}%
  \BibitemOpen
  \bibfield  {author} {\bibinfo {author} {\bibfnamefont {A.}~\bibnamefont
  {Escriv\`a}}, \bibinfo {author} {\bibfnamefont {C.}~\bibnamefont {Germani}},
  \ and\ \bibinfo {author} {\bibfnamefont {R.~K.}\ \bibnamefont {Sheth}},\
  }\href {\doibase 10.1103/PhysRevD.101.044022} {\bibfield  {journal} {\bibinfo
   {journal} {Phys. Rev. D}\ }\textbf {\bibinfo {volume} {101}},\ \bibinfo
  {pages} {044022} (\bibinfo {year} {2020})},\ \Eprint
  {http://arxiv.org/abs/1907.13311} {arXiv:1907.13311 [gr-qc]} \BibitemShut
  {NoStop}%
\bibitem [{\citenamefont {Escriv\`a}(2022)}]{Escriva:2021aeh}%
  \BibitemOpen
  \bibfield  {author} {\bibinfo {author} {\bibfnamefont {A.}~\bibnamefont
  {Escriv\`a}},\ }\href {\doibase 10.3390/universe8020066} {\bibfield
  {journal} {\bibinfo  {journal} {Universe}\ }\textbf {\bibinfo {volume} {8}},\
  \bibinfo {pages} {66} (\bibinfo {year} {2022})},\ \Eprint
  {http://arxiv.org/abs/2111.12693} {arXiv:2111.12693 [gr-qc]} \BibitemShut
  {NoStop}%
\bibitem [{\citenamefont {Jedamzik}(1995)}]{Jedamzik:1994nr}%
  \BibitemOpen
  \bibfield  {author} {\bibinfo {author} {\bibfnamefont {K.}~\bibnamefont
  {Jedamzik}},\ }\href {\doibase 10.1086/175936} {\bibfield  {journal}
  {\bibinfo  {journal} {Astrophys. J.}\ }\textbf {\bibinfo {volume} {448}},\
  \bibinfo {pages} {1} (\bibinfo {year} {1995})},\ \Eprint
  {http://arxiv.org/abs/astro-ph/9408080} {arXiv:astro-ph/9408080} \BibitemShut
  {NoStop}%
\bibitem [{\citenamefont {Kushwaha}\ and\ \citenamefont
  {Suyama}(2025)}]{Kushwaha:2025zpz}%
  \BibitemOpen
  \bibfield  {author} {\bibinfo {author} {\bibfnamefont {A.}~\bibnamefont
  {Kushwaha}}\ and\ \bibinfo {author} {\bibfnamefont {T.}~\bibnamefont
  {Suyama}},\ }\href@noop {} {\  (\bibinfo {year} {2025})},\ \Eprint
  {http://arxiv.org/abs/2509.25871} {arXiv:2509.25871 [astro-ph.CO]}
  \BibitemShut {NoStop}%
\bibitem [{\citenamefont {Auclair}\ and\ \citenamefont
  {Vennin}(2021)}]{Auclair:2020csm}%
  \BibitemOpen
  \bibfield  {author} {\bibinfo {author} {\bibfnamefont {P.}~\bibnamefont
  {Auclair}}\ and\ \bibinfo {author} {\bibfnamefont {V.}~\bibnamefont
  {Vennin}},\ }\href@noop {} {\bibfield  {journal} {\bibinfo  {journal} {JCAP}\
  }\textbf {\bibinfo {volume} {02}},\ \bibinfo {pages} {038} (\bibinfo {year}
  {2021})},\ \Eprint {http://arxiv.org/abs/2011.05633} {arXiv:2011.05633}
  \BibitemShut {NoStop}%
\bibitem [{\citenamefont {Moradinezhad~Dizgah}\ \emph
  {et~al.}(2019)\citenamefont {Moradinezhad~Dizgah}, \citenamefont
  {Franciolini},\ and\ \citenamefont {Riotto}}]{MoradinezhadDizgah:2019wjf}%
  \BibitemOpen
  \bibfield  {author} {\bibinfo {author} {\bibfnamefont {A.}~\bibnamefont
  {Moradinezhad~Dizgah}}, \bibinfo {author} {\bibfnamefont {G.}~\bibnamefont
  {Franciolini}}, \ and\ \bibinfo {author} {\bibfnamefont {A.}~\bibnamefont
  {Riotto}},\ }\href {\doibase 10.1088/1475-7516/2019/11/001} {\bibfield
  {journal} {\bibinfo  {journal} {JCAP}\ }\textbf {\bibinfo {volume} {11}},\
  \bibinfo {pages} {001} (\bibinfo {year} {2019})},\ \Eprint
  {http://arxiv.org/abs/1906.08978} {arXiv:1906.08978 [astro-ph.CO]}
  \BibitemShut {NoStop}%
\bibitem [{\citenamefont {De~Luca}\ \emph {et~al.}(2020)\citenamefont
  {De~Luca}, \citenamefont {Franciolini},\ and\ \citenamefont
  {Riotto}}]{DeLuca:2020ioi}%
  \BibitemOpen
  \bibfield  {author} {\bibinfo {author} {\bibfnamefont {V.}~\bibnamefont
  {De~Luca}}, \bibinfo {author} {\bibfnamefont {G.}~\bibnamefont
  {Franciolini}}, \ and\ \bibinfo {author} {\bibfnamefont {A.}~\bibnamefont
  {Riotto}},\ }\href {\doibase 10.1016/j.physletb.2020.135550} {\bibfield
  {journal} {\bibinfo  {journal} {Phys. Lett. B}\ }\textbf {\bibinfo {volume}
  {807}},\ \bibinfo {pages} {135550} (\bibinfo {year} {2020})},\ \Eprint
  {http://arxiv.org/abs/2001.04371} {arXiv:2001.04371 [astro-ph.CO]}
  \BibitemShut {NoStop}%
\bibitem [{\citenamefont {Auclair}\ and\ \citenamefont
  {Blachier}(2024)}]{Auclair:2024jwj}%
  \BibitemOpen
  \bibfield  {author} {\bibinfo {author} {\bibfnamefont {P.}~\bibnamefont
  {Auclair}}\ and\ \bibinfo {author} {\bibfnamefont {B.}~\bibnamefont
  {Blachier}},\ }\href {\doibase 10.1103/PhysRevD.109.123538} {\bibfield
  {journal} {\bibinfo  {journal} {Phys. Rev. D}\ }\textbf {\bibinfo {volume}
  {109}},\ \bibinfo {pages} {123538} (\bibinfo {year} {2024})},\ \Eprint
  {http://arxiv.org/abs/2402.00600} {arXiv:2402.00600 [astro-ph.CO]}
  \BibitemShut {NoStop}%
\bibitem [{\citenamefont {Dizon}(2025)}]{Dizon:2025siw}%
  \BibitemOpen
  \bibfield  {author} {\bibinfo {author} {\bibfnamefont {G.~L.}\ \bibnamefont
  {Dizon}},\ }\href {\doibase 10.1103/cwsc-p4v8} {\bibfield  {journal}
  {\bibinfo  {journal} {Phys. Rev. D}\ }\textbf {\bibinfo {volume} {112}},\
  \bibinfo {pages} {123024} (\bibinfo {year} {2025})},\ \Eprint
  {http://arxiv.org/abs/2506.06648} {arXiv:2506.06648 [astro-ph.CO]}
  \BibitemShut {NoStop}%
\bibitem [{\citenamefont {Kameli}\ and\ \citenamefont
  {Erfani}(2025)}]{Kameli:2025qzp}%
  \BibitemOpen
  \bibfield  {author} {\bibinfo {author} {\bibfnamefont {H.}~\bibnamefont
  {Kameli}}\ and\ \bibinfo {author} {\bibfnamefont {E.}~\bibnamefont
  {Erfani}},\ }\href@noop {} {\  (\bibinfo {year} {2025})},\ \Eprint
  {http://arxiv.org/abs/2508.01896} {arXiv:2508.01896 [astro-ph.CO]}
  \BibitemShut {NoStop}%
\bibitem [{\citenamefont {Akrami}\ \emph {et~al.}(2020)\citenamefont {Akrami}
  \emph {et~al.}}]{Planck:2019kim}%
  \BibitemOpen
  \bibfield  {author} {\bibinfo {author} {\bibfnamefont {Y.}~\bibnamefont
  {Akrami}} \emph {et~al.} (\bibinfo {collaboration} {Planck}),\ }\href
  {\doibase 10.1051/0004-6361/201935891} {\bibfield  {journal} {\bibinfo
  {journal} {Astron. Astrophys.}\ }\textbf {\bibinfo {volume} {641}},\ \bibinfo
  {pages} {A9} (\bibinfo {year} {2020})},\ \Eprint
  {http://arxiv.org/abs/1905.05697} {arXiv:1905.05697 [astro-ph.CO]}
  \BibitemShut {NoStop}%
\bibitem [{\citenamefont {Saito}\ and\ \citenamefont
  {Tokeshi}(2025)}]{Saito:2025sny}%
  \BibitemOpen
  \bibfield  {author} {\bibinfo {author} {\bibfnamefont {D.}~\bibnamefont
  {Saito}}\ and\ \bibinfo {author} {\bibfnamefont {K.}~\bibnamefont
  {Tokeshi}},\ }\href@noop {} {\  (\bibinfo {year} {2025})},\ \Eprint
  {http://arxiv.org/abs/2512.22075} {arXiv:2512.22075 [astro-ph.CO]}
  \BibitemShut {NoStop}%
\bibitem [{\citenamefont {Paranjape}\ \emph {et~al.}(2012)\citenamefont
  {Paranjape}, \citenamefont {Lam},\ and\ \citenamefont
  {Sheth}}]{Paranjape:2011wa}%
  \BibitemOpen
  \bibfield  {author} {\bibinfo {author} {\bibfnamefont {A.}~\bibnamefont
  {Paranjape}}, \bibinfo {author} {\bibfnamefont {T.~Y.}\ \bibnamefont {Lam}},
  \ and\ \bibinfo {author} {\bibfnamefont {R.~K.}\ \bibnamefont {Sheth}},\
  }\href {\doibase 10.1111/j.1365-2966.2011.20128.x} {\bibfield  {journal}
  {\bibinfo  {journal} {Mon. Not. Roy. Astron. Soc.}\ }\textbf {\bibinfo
  {volume} {420}},\ \bibinfo {pages} {1429} (\bibinfo {year} {2012})},\ \Eprint
  {http://arxiv.org/abs/1105.1990} {arXiv:1105.1990 [astro-ph.CO]} \BibitemShut
  {NoStop}%
\bibitem [{\citenamefont {Musso}\ and\ \citenamefont
  {Paranjape}(2012)}]{Musso:2011ck}%
  \BibitemOpen
  \bibfield  {author} {\bibinfo {author} {\bibfnamefont {M.}~\bibnamefont
  {Musso}}\ and\ \bibinfo {author} {\bibfnamefont {A.}~\bibnamefont
  {Paranjape}},\ }\href {\doibase 10.1111/j.1365-2966.2011.20040.x} {\bibfield
  {journal} {\bibinfo  {journal} {Mon. Not. Roy. Astron. Soc.}\ }\textbf
  {\bibinfo {volume} {420}},\ \bibinfo {pages} {369} (\bibinfo {year}
  {2012})},\ \Eprint {http://arxiv.org/abs/1108.0565} {arXiv:1108.0565
  [astro-ph.CO]} \BibitemShut {NoStop}%
\bibitem [{\citenamefont {Musso}\ and\ \citenamefont
  {Sheth}(2014)}]{Musso:2013pha}%
  \BibitemOpen
  \bibfield  {author} {\bibinfo {author} {\bibfnamefont {M.}~\bibnamefont
  {Musso}}\ and\ \bibinfo {author} {\bibfnamefont {R.~K.}\ \bibnamefont
  {Sheth}},\ }\href@noop {} {\bibfield  {journal} {\bibinfo  {journal} {Mon.
  Not. Roy. Astron. Soc.}\ }\textbf {\bibinfo {volume} {438}},\ \bibinfo
  {pages} {2683} (\bibinfo {year} {2014})},\ \Eprint
  {http://arxiv.org/abs/1306.0551} {arXiv:1306.0551 [astro-ph.CO]} \BibitemShut
  {NoStop}%
\bibitem [{\citenamefont {Nikakhtar}\ \emph {et~al.}(2018)\citenamefont
  {Nikakhtar}, \citenamefont {Ayromlou}, \citenamefont {Baghram}, \citenamefont
  {Rahvar}, \citenamefont {Rahimi~Tabar},\ and\ \citenamefont
  {Sheth}}]{Nikakhtar:2018qqg}%
  \BibitemOpen
  \bibfield  {author} {\bibinfo {author} {\bibfnamefont {F.}~\bibnamefont
  {Nikakhtar}}, \bibinfo {author} {\bibfnamefont {M.}~\bibnamefont {Ayromlou}},
  \bibinfo {author} {\bibfnamefont {S.}~\bibnamefont {Baghram}}, \bibinfo
  {author} {\bibfnamefont {S.}~\bibnamefont {Rahvar}}, \bibinfo {author}
  {\bibfnamefont {M.~R.}\ \bibnamefont {Rahimi~Tabar}}, \ and\ \bibinfo
  {author} {\bibfnamefont {R.~K.}\ \bibnamefont {Sheth}},\ }\href {\doibase
  10.1093/mnras/sty1415} {\bibfield  {journal} {\bibinfo  {journal} {Mon. Not.
  Roy. Astron. Soc.}\ }\textbf {\bibinfo {volume} {478}},\ \bibinfo {pages}
  {5296} (\bibinfo {year} {2018})},\ \Eprint {http://arxiv.org/abs/1802.04207}
  {arXiv:1802.04207 [astro-ph.CO]} \BibitemShut {NoStop}%
\bibitem [{\citenamefont {Young}\ \emph {et~al.}(2014)\citenamefont {Young},
  \citenamefont {Byrnes},\ and\ \citenamefont {Sasaki}}]{Young:2014ana}%
  \BibitemOpen
  \bibfield  {author} {\bibinfo {author} {\bibfnamefont {S.}~\bibnamefont
  {Young}}, \bibinfo {author} {\bibfnamefont {C.~T.}\ \bibnamefont {Byrnes}}, \
  and\ \bibinfo {author} {\bibfnamefont {M.}~\bibnamefont {Sasaki}},\ }\href
  {\doibase 10.1088/1475-7516/2014/07/045} {\bibfield  {journal} {\bibinfo
  {journal} {JCAP}\ }\textbf {\bibinfo {volume} {07}},\ \bibinfo {pages} {045}
  (\bibinfo {year} {2014})},\ \Eprint {http://arxiv.org/abs/1405.7023}
  {arXiv:1405.7023 [gr-qc]} \BibitemShut {NoStop}%
\bibitem [{\citenamefont {Germani}\ and\ \citenamefont
  {Musco}(2019)}]{Germani:2018jgr}%
  \BibitemOpen
  \bibfield  {author} {\bibinfo {author} {\bibfnamefont {C.}~\bibnamefont
  {Germani}}\ and\ \bibinfo {author} {\bibfnamefont {I.}~\bibnamefont
  {Musco}},\ }\href {\doibase 10.1103/PhysRevLett.122.141302} {\bibfield
  {journal} {\bibinfo  {journal} {Phys. Rev. Lett.}\ }\textbf {\bibinfo
  {volume} {122}},\ \bibinfo {pages} {141302} (\bibinfo {year} {2019})},\
  \Eprint {http://arxiv.org/abs/1805.04087} {arXiv:1805.04087 [astro-ph.CO]}
  \BibitemShut {NoStop}%
\bibitem [{\citenamefont {Musco}\ \emph {et~al.}(2021)\citenamefont {Musco},
  \citenamefont {De~Luca}, \citenamefont {Franciolini},\ and\ \citenamefont
  {Riotto}}]{Musco:2020jjb}%
  \BibitemOpen
  \bibfield  {author} {\bibinfo {author} {\bibfnamefont {I.}~\bibnamefont
  {Musco}}, \bibinfo {author} {\bibfnamefont {V.}~\bibnamefont {De~Luca}},
  \bibinfo {author} {\bibfnamefont {G.}~\bibnamefont {Franciolini}}, \ and\
  \bibinfo {author} {\bibfnamefont {A.}~\bibnamefont {Riotto}},\ }\href
  {\doibase 10.1103/PhysRevD.103.063538} {\bibfield  {journal} {\bibinfo
  {journal} {Phys. Rev. D}\ }\textbf {\bibinfo {volume} {103}},\ \bibinfo
  {pages} {063538} (\bibinfo {year} {2021})},\ \Eprint
  {http://arxiv.org/abs/2011.03014} {arXiv:2011.03014 [astro-ph.CO]}
  \BibitemShut {NoStop}%
\bibitem [{\citenamefont {Shimada}\ \emph {et~al.}(2025)\citenamefont
  {Shimada}, \citenamefont {Escriv{\'a}}, \citenamefont {Saito}, \citenamefont
  {Uehara},\ and\ \citenamefont {Yoo}}]{Shimada:2024eec}%
  \BibitemOpen
  \bibfield  {author} {\bibinfo {author} {\bibfnamefont {M.}~\bibnamefont
  {Shimada}}, \bibinfo {author} {\bibfnamefont {A.}~\bibnamefont
  {Escriv{\'a}}}, \bibinfo {author} {\bibfnamefont {D.}~\bibnamefont {Saito}},
  \bibinfo {author} {\bibfnamefont {K.}~\bibnamefont {Uehara}}, \ and\ \bibinfo
  {author} {\bibfnamefont {C.-M.}\ \bibnamefont {Yoo}},\ }\href {\doibase
  10.1088/1475-7516/2025/02/018} {\bibfield  {journal} {\bibinfo  {journal}
  {JCAP}\ }\textbf {\bibinfo {volume} {02}},\ \bibinfo {pages} {018} (\bibinfo
  {year} {2025})},\ \Eprint {http://arxiv.org/abs/2411.07648} {arXiv:2411.07648
  [gr-qc]} \BibitemShut {NoStop}%
\bibitem [{\citenamefont {Inui}\ \emph {et~al.}(2025)\citenamefont {Inui},
  \citenamefont {Joana}, \citenamefont {Motohashi}, \citenamefont {Pi},
  \citenamefont {Tada},\ and\ \citenamefont {Yokoyama}}]{Inui:2024fgk}%
  \BibitemOpen
  \bibfield  {author} {\bibinfo {author} {\bibfnamefont {R.}~\bibnamefont
  {Inui}}, \bibinfo {author} {\bibfnamefont {C.}~\bibnamefont {Joana}},
  \bibinfo {author} {\bibfnamefont {H.}~\bibnamefont {Motohashi}}, \bibinfo
  {author} {\bibfnamefont {S.}~\bibnamefont {Pi}}, \bibinfo {author}
  {\bibfnamefont {Y.}~\bibnamefont {Tada}}, \ and\ \bibinfo {author}
  {\bibfnamefont {S.}~\bibnamefont {Yokoyama}},\ }\href {\doibase
  10.1088/1475-7516/2025/03/021} {\bibfield  {journal} {\bibinfo  {journal}
  {JCAP}\ }\textbf {\bibinfo {volume} {03}},\ \bibinfo {pages} {021} (\bibinfo
  {year} {2025})},\ \Eprint {http://arxiv.org/abs/2411.07647} {arXiv:2411.07647
  [astro-ph.CO]} \BibitemShut {NoStop}%
\bibitem [{\citenamefont {Escriv{\`a}}(2025{\natexlab{a}})}]{Escriva:2025eqc}%
  \BibitemOpen
  \bibfield  {author} {\bibinfo {author} {\bibfnamefont {A.}~\bibnamefont
  {Escriv{\`a}}},\ }\href {\doibase 10.1016/j.dark.2025.102177} {\bibfield
  {journal} {\bibinfo  {journal} {Phys. Dark Univ.}\ }\textbf {\bibinfo
  {volume} {50}},\ \bibinfo {pages} {102177} (\bibinfo {year}
  {2025}{\natexlab{a}})},\ \Eprint {http://arxiv.org/abs/2504.05813}
  {arXiv:2504.05813 [astro-ph.CO]} \BibitemShut {NoStop}%
\bibitem [{\citenamefont {Escriv{\`a}}(2025{\natexlab{b}})}]{Escriva:2025rja}%
  \BibitemOpen
  \bibfield  {author} {\bibinfo {author} {\bibfnamefont {A.}~\bibnamefont
  {Escriv{\`a}}},\ }\href {\doibase 10.1103/mq67-bbvj} {\bibfield  {journal}
  {\bibinfo  {journal} {Phys. Rev. D}\ }\textbf {\bibinfo {volume} {112}},\
  \bibinfo {pages} {103527} (\bibinfo {year} {2025}{\natexlab{b}})},\ \Eprint
  {http://arxiv.org/abs/2504.05814} {arXiv:2504.05814 [astro-ph.CO]}
  \BibitemShut {NoStop}%
\bibitem [{\citenamefont {Sureda}\ \emph {et~al.}(2021)\citenamefont {Sureda},
  \citenamefont {Magana}, \citenamefont {Araya},\ and\ \citenamefont
  {Padilla}}]{Sureda:2020vgi}%
  \BibitemOpen
  \bibfield  {author} {\bibinfo {author} {\bibfnamefont {J.}~\bibnamefont
  {Sureda}}, \bibinfo {author} {\bibfnamefont {J.}~\bibnamefont {Magana}},
  \bibinfo {author} {\bibfnamefont {I.~J.}\ \bibnamefont {Araya}}, \ and\
  \bibinfo {author} {\bibfnamefont {N.~D.}\ \bibnamefont {Padilla}},\ }\href
  {\doibase 10.1093/mnras/stab2450} {\bibfield  {journal} {\bibinfo  {journal}
  {Mon. Not. Roy. Astron. Soc.}\ }\textbf {\bibinfo {volume} {507}},\ \bibinfo
  {pages} {4804} (\bibinfo {year} {2021})},\ \Eprint
  {http://arxiv.org/abs/2008.09683} {arXiv:2008.09683 [astro-ph.CO]}
  \BibitemShut {NoStop}%
\bibitem [{\citenamefont {Tuckwell}\ and\ \citenamefont
  {Wan}(1984)}]{Tuckwell_Wan_1984}%
  \BibitemOpen
  \bibfield  {author} {\bibinfo {author} {\bibfnamefont {H.~C.}\ \bibnamefont
  {Tuckwell}}\ and\ \bibinfo {author} {\bibfnamefont {F.~Y.~M.}\ \bibnamefont
  {Wan}},\ }\href {\doibase 10.2307/3213688} {\bibfield  {journal} {\bibinfo
  {journal} {Journal of Applied Probability}\ }\textbf {\bibinfo {volume}
  {21}},\ \bibinfo {pages} {695–709} (\bibinfo {year} {1984})}\BibitemShut
  {NoStop}%
\bibitem [{\citenamefont {Buonocore}\ \emph {et~al.}(1987)\citenamefont
  {Buonocore}, \citenamefont {Nobile},\ and\ \citenamefont
  {Ricciardi}}]{Buonocore1987ANI}%
  \BibitemOpen
  \bibfield  {author} {\bibinfo {author} {\bibfnamefont {A.}~\bibnamefont
  {Buonocore}}, \bibinfo {author} {\bibfnamefont {A.~G.}\ \bibnamefont
  {Nobile}}, \ and\ \bibinfo {author} {\bibfnamefont {L.~M.}\ \bibnamefont
  {Ricciardi}},\ }\href {https://api.semanticscholar.org/CorpusID:123146466}
  {\bibfield  {journal} {\bibinfo  {journal} {Advances in Applied Probability}\
  }\textbf {\bibinfo {volume} {19}},\ \bibinfo {pages} {784 } (\bibinfo {year}
  {1987})}\BibitemShut {NoStop}%
\bibitem [{\citenamefont {Zhang}\ and\ \citenamefont
  {Hui}(2006)}]{Zhang:2005ar}%
  \BibitemOpen
  \bibfield  {author} {\bibinfo {author} {\bibfnamefont {J.}~\bibnamefont
  {Zhang}}\ and\ \bibinfo {author} {\bibfnamefont {L.}~\bibnamefont {Hui}},\
  }\href {\doibase 10.1086/499802} {\bibfield  {journal} {\bibinfo  {journal}
  {Astrophys. J.}\ }\textbf {\bibinfo {volume} {641}},\ \bibinfo {pages} {641}
  (\bibinfo {year} {2006})},\ \Eprint {http://arxiv.org/abs/astro-ph/0508384}
  {arXiv:astro-ph/0508384} \BibitemShut {NoStop}%
\bibitem [{\citenamefont {Molini}\ \emph {et~al.}(2011)\citenamefont {Molini},
  \citenamefont {Talkner}, \citenamefont {Katul},\ and\ \citenamefont
  {Porporato}}]{Molini2011}%
  \BibitemOpen
  \bibfield  {author} {\bibinfo {author} {\bibfnamefont {A.}~\bibnamefont
  {Molini}}, \bibinfo {author} {\bibfnamefont {P.}~\bibnamefont {Talkner}},
  \bibinfo {author} {\bibfnamefont {G.}~\bibnamefont {Katul}}, \ and\ \bibinfo
  {author} {\bibfnamefont {A.}~\bibnamefont {Porporato}},\ }\href {\doibase
  https://doi.org/10.1016/j.physa.2011.01.024} {\bibfield  {journal} {\bibinfo
  {journal} {Physica A: Statistical Mechanics and its Applications}\ }\textbf
  {\bibinfo {volume} {390}},\ \bibinfo {pages} {1841} (\bibinfo {year}
  {2011})}\BibitemShut {NoStop}%
\bibitem [{\citenamefont {Giorno}\ \emph {et~al.}(1989)\citenamefont {Giorno},
  \citenamefont {Nobile}, \citenamefont {Ricciardi},\ and\ \citenamefont
  {Sato}}]{Giorno:1989}%
  \BibitemOpen
  \bibfield  {author} {\bibinfo {author} {\bibfnamefont {V.}~\bibnamefont
  {Giorno}}, \bibinfo {author} {\bibfnamefont {A.~G.}\ \bibnamefont {Nobile}},
  \bibinfo {author} {\bibfnamefont {L.~M.}\ \bibnamefont {Ricciardi}}, \ and\
  \bibinfo {author} {\bibfnamefont {S.}~\bibnamefont {Sato}},\ }\href
  {http://www.jstor.org/stable/1427196} {\bibfield  {journal} {\bibinfo
  {journal} {Advances in Applied Probability}\ }\textbf {\bibinfo {volume}
  {21}},\ \bibinfo {pages} {20} (\bibinfo {year} {1989})}\BibitemShut {NoStop}%
\bibitem [{\citenamefont {Strassen}(1969)}]{strassen1969gaussian}%
  \BibitemOpen
  \bibfield  {author} {\bibinfo {author} {\bibfnamefont {V.}~\bibnamefont
  {Strassen}},\ }\href@noop {} {\bibfield  {journal} {\bibinfo  {journal}
  {Numerische mathematik}\ }\textbf {\bibinfo {volume} {13}},\ \bibinfo {pages}
  {354} (\bibinfo {year} {1969})}\BibitemShut {NoStop}%
\bibitem [{\citenamefont {Press}(2007)}]{press2007numerical}%
  \BibitemOpen
  \bibfield  {author} {\bibinfo {author} {\bibfnamefont {W.~H.}\ \bibnamefont
  {Press}},\ }\href@noop {} {\emph {\bibinfo {title} {Numerical recipes 3rd
  edition: The art of scientific computing}}}\ (\bibinfo  {publisher}
  {Cambridge university press},\ \bibinfo {year} {2007})\BibitemShut {NoStop}%
\bibitem [{\citenamefont {Alman}\ \emph {et~al.}(2025)\citenamefont {Alman},
  \citenamefont {Duan}, \citenamefont {Williams}, \citenamefont {Xu},
  \citenamefont {Xu},\ and\ \citenamefont
  {Zhou}}]{doi:10.1137/1.9781611978322.63}%
  \BibitemOpen
  \bibfield  {author} {\bibinfo {author} {\bibfnamefont {J.}~\bibnamefont
  {Alman}}, \bibinfo {author} {\bibfnamefont {R.}~\bibnamefont {Duan}},
  \bibinfo {author} {\bibfnamefont {V.~V.}\ \bibnamefont {Williams}}, \bibinfo
  {author} {\bibfnamefont {Y.}~\bibnamefont {Xu}}, \bibinfo {author}
  {\bibfnamefont {Z.}~\bibnamefont {Xu}}, \ and\ \bibinfo {author}
  {\bibfnamefont {R.}~\bibnamefont {Zhou}},\ }\enquote {\bibinfo {title} {More
  asymmetry yields faster matrix multiplication},}\ in\ \href {\doibase
  10.1137/1.9781611978322.63} {\emph {\bibinfo {booktitle} {Proceedings of the
  2025 Annual ACM-SIAM Symposium on Discrete Algorithms (SODA)}}}\ (\bibinfo
  {year} {2025})\ pp.\ \bibinfo {pages} {2005--2039}\BibitemShut {NoStop}%
\bibitem [{\citenamefont {Choptuik}(1993)}]{Choptuik:1992jv}%
  \BibitemOpen
  \bibfield  {author} {\bibinfo {author} {\bibfnamefont {M.~W.}\ \bibnamefont
  {Choptuik}},\ }\href {\doibase 10.1103/PhysRevLett.70.9} {\bibfield
  {journal} {\bibinfo  {journal} {Phys. Rev. Lett.}\ }\textbf {\bibinfo
  {volume} {70}},\ \bibinfo {pages} {9} (\bibinfo {year} {1993})}\BibitemShut
  {NoStop}%
\bibitem [{\citenamefont {Niemeyer}\ and\ \citenamefont
  {Jedamzik}(1998)}]{Niemeyer:1997mt}%
  \BibitemOpen
  \bibfield  {author} {\bibinfo {author} {\bibfnamefont {J.~C.}\ \bibnamefont
  {Niemeyer}}\ and\ \bibinfo {author} {\bibfnamefont {K.}~\bibnamefont
  {Jedamzik}},\ }\href {\doibase 10.1103/PhysRevLett.80.5481} {\bibfield
  {journal} {\bibinfo  {journal} {Phys. Rev. Lett.}\ }\textbf {\bibinfo
  {volume} {80}},\ \bibinfo {pages} {5481} (\bibinfo {year} {1998})},\ \Eprint
  {http://arxiv.org/abs/astro-ph/9709072} {arXiv:astro-ph/9709072} \BibitemShut
  {NoStop}%
\bibitem [{\citenamefont {Musco}\ \emph {et~al.}(2009)\citenamefont {Musco},
  \citenamefont {Miller},\ and\ \citenamefont {Polnarev}}]{Musco:2008hv}%
  \BibitemOpen
  \bibfield  {author} {\bibinfo {author} {\bibfnamefont {I.}~\bibnamefont
  {Musco}}, \bibinfo {author} {\bibfnamefont {J.~C.}\ \bibnamefont {Miller}}, \
  and\ \bibinfo {author} {\bibfnamefont {A.~G.}\ \bibnamefont {Polnarev}},\
  }\href {\doibase 10.1088/0264-9381/26/23/235001} {\bibfield  {journal}
  {\bibinfo  {journal} {Class. Quant. Grav.}\ }\textbf {\bibinfo {volume}
  {26}},\ \bibinfo {pages} {235001} (\bibinfo {year} {2009})},\ \Eprint
  {http://arxiv.org/abs/0811.1452} {arXiv:0811.1452 [gr-qc]} \BibitemShut
  {NoStop}%
\bibitem [{\citenamefont {Musco}\ and\ \citenamefont
  {Miller}(2013)}]{Musco:2012au}%
  \BibitemOpen
  \bibfield  {author} {\bibinfo {author} {\bibfnamefont {I.}~\bibnamefont
  {Musco}}\ and\ \bibinfo {author} {\bibfnamefont {J.~C.}\ \bibnamefont
  {Miller}},\ }\href {\doibase 10.1088/0264-9381/30/14/145009} {\bibfield
  {journal} {\bibinfo  {journal} {Class. Quant. Grav.}\ }\textbf {\bibinfo
  {volume} {30}},\ \bibinfo {pages} {145009} (\bibinfo {year} {2013})},\
  \Eprint {http://arxiv.org/abs/1201.2379} {arXiv:1201.2379 [gr-qc]}
  \BibitemShut {NoStop}%
\bibitem [{\citenamefont {Harada}\ \emph {et~al.}(2015)\citenamefont {Harada},
  \citenamefont {Yoo}, \citenamefont {Nakama},\ and\ \citenamefont
  {Koga}}]{Harada:2015yda}%
  \BibitemOpen
  \bibfield  {author} {\bibinfo {author} {\bibfnamefont {T.}~\bibnamefont
  {Harada}}, \bibinfo {author} {\bibfnamefont {C.-M.}\ \bibnamefont {Yoo}},
  \bibinfo {author} {\bibfnamefont {T.}~\bibnamefont {Nakama}}, \ and\ \bibinfo
  {author} {\bibfnamefont {Y.}~\bibnamefont {Koga}},\ }\href {\doibase
  10.1103/PhysRevD.91.084057} {\bibfield  {journal} {\bibinfo  {journal} {Phys.
  Rev. D}\ }\textbf {\bibinfo {volume} {91}},\ \bibinfo {pages} {084057}
  (\bibinfo {year} {2015})},\ \Eprint {http://arxiv.org/abs/1503.03934}
  {arXiv:1503.03934 [gr-qc]} \BibitemShut {NoStop}%
\bibitem [{\citenamefont {Yokoyama}(1998)}]{Yokoyama:1998xd}%
  \BibitemOpen
  \bibfield  {author} {\bibinfo {author} {\bibfnamefont {J.}~\bibnamefont
  {Yokoyama}},\ }\href {\doibase 10.1103/PhysRevD.58.107502} {\bibfield
  {journal} {\bibinfo  {journal} {Phys. Rev. D}\ }\textbf {\bibinfo {volume}
  {58}},\ \bibinfo {pages} {107502} (\bibinfo {year} {1998})},\ \Eprint
  {http://arxiv.org/abs/gr-qc/9804041} {arXiv:gr-qc/9804041} \BibitemShut
  {NoStop}%
\bibitem [{\citenamefont {Green}\ and\ \citenamefont
  {Liddle}(1999)}]{Green:1999xm}%
  \BibitemOpen
  \bibfield  {author} {\bibinfo {author} {\bibfnamefont {A.~M.}\ \bibnamefont
  {Green}}\ and\ \bibinfo {author} {\bibfnamefont {A.~R.}\ \bibnamefont
  {Liddle}},\ }\href {\doibase 10.1103/PhysRevD.60.063509} {\bibfield
  {journal} {\bibinfo  {journal} {Phys. Rev. D}\ }\textbf {\bibinfo {volume}
  {60}},\ \bibinfo {pages} {063509} (\bibinfo {year} {1999})},\ \Eprint
  {http://arxiv.org/abs/astro-ph/9901268} {arXiv:astro-ph/9901268} \BibitemShut
  {NoStop}%
\bibitem [{\citenamefont {Germani}\ and\ \citenamefont
  {Sheth}(2023)}]{Germani:2023ojx}%
  \BibitemOpen
  \bibfield  {author} {\bibinfo {author} {\bibfnamefont {C.}~\bibnamefont
  {Germani}}\ and\ \bibinfo {author} {\bibfnamefont {R.~K.}\ \bibnamefont
  {Sheth}},\ }\href {\doibase 10.3390/universe9090421} {\bibfield  {journal}
  {\bibinfo  {journal} {Universe}\ }\textbf {\bibinfo {volume} {9}},\ \bibinfo
  {pages} {421} (\bibinfo {year} {2023})},\ \Eprint
  {http://arxiv.org/abs/2308.02971} {arXiv:2308.02971 [astro-ph.CO]}
  \BibitemShut {NoStop}%
\bibitem [{\citenamefont {Pi}\ and\ \citenamefont {Wang}(2023)}]{Pi:2022zxs}%
  \BibitemOpen
  \bibfield  {author} {\bibinfo {author} {\bibfnamefont {S.}~\bibnamefont
  {Pi}}\ and\ \bibinfo {author} {\bibfnamefont {J.}~\bibnamefont {Wang}},\
  }\href {\doibase 10.1088/1475-7516/2023/06/018} {\bibfield  {journal}
  {\bibinfo  {journal} {JCAP}\ }\textbf {\bibinfo {volume} {06}},\ \bibinfo
  {pages} {018} (\bibinfo {year} {2023})},\ \Eprint
  {http://arxiv.org/abs/2209.14183} {arXiv:2209.14183 [astro-ph.CO]}
  \BibitemShut {NoStop}%
\bibitem [{\citenamefont {Dom{\`e}nech}\ \emph {et~al.}(2024)\citenamefont
  {Dom{\`e}nech}, \citenamefont {Vargas},\ and\ \citenamefont
  {Vargas}}]{Domenech:2023dxx}%
  \BibitemOpen
  \bibfield  {author} {\bibinfo {author} {\bibfnamefont {G.}~\bibnamefont
  {Dom{\`e}nech}}, \bibinfo {author} {\bibfnamefont {G.}~\bibnamefont
  {Vargas}}, \ and\ \bibinfo {author} {\bibfnamefont {T.}~\bibnamefont
  {Vargas}},\ }\href {\doibase 10.1088/1475-7516/2024/03/002} {\bibfield
  {journal} {\bibinfo  {journal} {JCAP}\ }\textbf {\bibinfo {volume} {03}},\
  \bibinfo {pages} {002} (\bibinfo {year} {2024})},\ \Eprint
  {http://arxiv.org/abs/2309.05750} {arXiv:2309.05750 [astro-ph.CO]}
  \BibitemShut {NoStop}%
\bibitem [{\citenamefont {Cielo}\ \emph {et~al.}(2025)\citenamefont {Cielo},
  \citenamefont {Mangano}, \citenamefont {Pisanti},\ and\ \citenamefont
  {Wands}}]{Cielo:2024poz}%
  \BibitemOpen
  \bibfield  {author} {\bibinfo {author} {\bibfnamefont {M.}~\bibnamefont
  {Cielo}}, \bibinfo {author} {\bibfnamefont {G.}~\bibnamefont {Mangano}},
  \bibinfo {author} {\bibfnamefont {O.}~\bibnamefont {Pisanti}}, \ and\
  \bibinfo {author} {\bibfnamefont {D.}~\bibnamefont {Wands}},\ }\href
  {\doibase 10.1088/1475-7516/2025/04/007} {\bibfield  {journal} {\bibinfo
  {journal} {JCAP}\ }\textbf {\bibinfo {volume} {04}},\ \bibinfo {pages} {007}
  (\bibinfo {year} {2025})},\ \Eprint {http://arxiv.org/abs/2410.22154}
  {arXiv:2410.22154 [astro-ph.CO]} \BibitemShut {NoStop}%
\bibitem [{\citenamefont {Briaud}\ \emph {et~al.}(2025)\citenamefont {Briaud},
  \citenamefont {Karam}, \citenamefont {Koivunen}, \citenamefont {Tomberg},
  \citenamefont {Veerm{\"a}e},\ and\ \citenamefont {Vennin}}]{Briaud:2025hra}%
  \BibitemOpen
  \bibfield  {author} {\bibinfo {author} {\bibfnamefont {V.}~\bibnamefont
  {Briaud}}, \bibinfo {author} {\bibfnamefont {A.}~\bibnamefont {Karam}},
  \bibinfo {author} {\bibfnamefont {N.}~\bibnamefont {Koivunen}}, \bibinfo
  {author} {\bibfnamefont {E.}~\bibnamefont {Tomberg}}, \bibinfo {author}
  {\bibfnamefont {H.}~\bibnamefont {Veerm{\"a}e}}, \ and\ \bibinfo {author}
  {\bibfnamefont {V.}~\bibnamefont {Vennin}},\ }\href {\doibase
  10.1088/1475-7516/2025/05/097} {\bibfield  {journal} {\bibinfo  {journal}
  {JCAP}\ }\textbf {\bibinfo {volume} {05}},\ \bibinfo {pages} {097} (\bibinfo
  {year} {2025})},\ \Eprint {http://arxiv.org/abs/2501.14681} {arXiv:2501.14681
  [astro-ph.CO]} \BibitemShut {NoStop}%
\bibitem [{\citenamefont {Saito}\ and\ \citenamefont
  {Yokoyama}(2010)}]{Saito:2009jt}%
  \BibitemOpen
  \bibfield  {author} {\bibinfo {author} {\bibfnamefont {R.}~\bibnamefont
  {Saito}}\ and\ \bibinfo {author} {\bibfnamefont {J.}~\bibnamefont
  {Yokoyama}},\ }\href {\doibase 10.1143/PTP.126.351} {\bibfield  {journal}
  {\bibinfo  {journal} {Prog. Theor. Phys.}\ }\textbf {\bibinfo {volume}
  {123}},\ \bibinfo {pages} {867} (\bibinfo {year} {2010})},\ \bibinfo {note}
  {[Erratum: Prog.Theor.Phys. 126, 351--352 (2011)]},\ \Eprint
  {http://arxiv.org/abs/0912.5317} {arXiv:0912.5317 [astro-ph.CO]} \BibitemShut
  {NoStop}%
\bibitem [{\citenamefont {Sugiyama}\ \emph {et~al.}(2021)\citenamefont
  {Sugiyama}, \citenamefont {Takhistov}, \citenamefont {Vitagliano},
  \citenamefont {Kusenko}, \citenamefont {Sasaki},\ and\ \citenamefont
  {Takada}}]{Sugiyama:2020roc}%
  \BibitemOpen
  \bibfield  {author} {\bibinfo {author} {\bibfnamefont {S.}~\bibnamefont
  {Sugiyama}}, \bibinfo {author} {\bibfnamefont {V.}~\bibnamefont {Takhistov}},
  \bibinfo {author} {\bibfnamefont {E.}~\bibnamefont {Vitagliano}}, \bibinfo
  {author} {\bibfnamefont {A.}~\bibnamefont {Kusenko}}, \bibinfo {author}
  {\bibfnamefont {M.}~\bibnamefont {Sasaki}}, \ and\ \bibinfo {author}
  {\bibfnamefont {M.}~\bibnamefont {Takada}},\ }\href {\doibase
  10.1016/j.physletb.2021.136097} {\bibfield  {journal} {\bibinfo  {journal}
  {Phys. Lett. B}\ }\textbf {\bibinfo {volume} {814}},\ \bibinfo {pages}
  {136097} (\bibinfo {year} {2021})},\ \Eprint
  {http://arxiv.org/abs/2010.02189} {arXiv:2010.02189 [astro-ph.CO]}
  \BibitemShut {NoStop}%
\bibitem [{\citenamefont {De~Luca}\ \emph {et~al.}(2021)\citenamefont
  {De~Luca}, \citenamefont {Franciolini},\ and\ \citenamefont
  {Riotto}}]{DeLuca:2020agl}%
  \BibitemOpen
  \bibfield  {author} {\bibinfo {author} {\bibfnamefont {V.}~\bibnamefont
  {De~Luca}}, \bibinfo {author} {\bibfnamefont {G.}~\bibnamefont
  {Franciolini}}, \ and\ \bibinfo {author} {\bibfnamefont {A.}~\bibnamefont
  {Riotto}},\ }\href {\doibase 10.1103/PhysRevLett.126.041303} {\bibfield
  {journal} {\bibinfo  {journal} {Phys. Rev. Lett.}\ }\textbf {\bibinfo
  {volume} {126}},\ \bibinfo {pages} {041303} (\bibinfo {year} {2021})},\
  \Eprint {http://arxiv.org/abs/2009.08268} {arXiv:2009.08268 [astro-ph.CO]}
  \BibitemShut {NoStop}%
\bibitem [{\citenamefont {Fumagalli}\ \emph {et~al.}(2025)\citenamefont
  {Fumagalli}, \citenamefont {Garriga}, \citenamefont {Germani},\ and\
  \citenamefont {Sheth}}]{Fumagalli:2024kgg}%
  \BibitemOpen
  \bibfield  {author} {\bibinfo {author} {\bibfnamefont {J.}~\bibnamefont
  {Fumagalli}}, \bibinfo {author} {\bibfnamefont {J.}~\bibnamefont {Garriga}},
  \bibinfo {author} {\bibfnamefont {C.}~\bibnamefont {Germani}}, \ and\
  \bibinfo {author} {\bibfnamefont {R.~K.}\ \bibnamefont {Sheth}},\ }\href
  {\doibase 10.1103/k75n-3qz4} {\bibfield  {journal} {\bibinfo  {journal}
  {Phys. Rev. D}\ }\textbf {\bibinfo {volume} {111}},\ \bibinfo {pages}
  {123518} (\bibinfo {year} {2025})},\ \Eprint
  {http://arxiv.org/abs/2412.07709} {arXiv:2412.07709 [astro-ph.CO]}
  \BibitemShut {NoStop}%
\bibitem [{\citenamefont {Young}(2019)}]{Young:2019osy}%
  \BibitemOpen
  \bibfield  {author} {\bibinfo {author} {\bibfnamefont {S.}~\bibnamefont
  {Young}},\ }\href {\doibase 10.1142/S0218271820300025} {\bibfield  {journal}
  {\bibinfo  {journal} {Int. J. Mod. Phys. D}\ }\textbf {\bibinfo {volume}
  {29}},\ \bibinfo {pages} {2030002} (\bibinfo {year} {2019})},\ \Eprint
  {http://arxiv.org/abs/1905.01230} {arXiv:1905.01230 [astro-ph.CO]}
  \BibitemShut {NoStop}%
\bibitem [{\citenamefont {Pi}\ and\ \citenamefont {Sasaki}(2020)}]{Pi:2020otn}%
  \BibitemOpen
  \bibfield  {author} {\bibinfo {author} {\bibfnamefont {S.}~\bibnamefont
  {Pi}}\ and\ \bibinfo {author} {\bibfnamefont {M.}~\bibnamefont {Sasaki}},\
  }\href {\doibase 10.1088/1475-7516/2020/09/037} {\bibfield  {journal}
  {\bibinfo  {journal} {JCAP}\ }\textbf {\bibinfo {volume} {09}},\ \bibinfo
  {pages} {037} (\bibinfo {year} {2020})},\ \Eprint
  {http://arxiv.org/abs/2005.12306} {arXiv:2005.12306 [gr-qc]} \BibitemShut
  {NoStop}%
\bibitem [{\citenamefont {Pi}\ \emph {et~al.}(2025)\citenamefont {Pi},
  \citenamefont {Sasaki}, \citenamefont {Takhistov},\ and\ \citenamefont
  {Wang}}]{Pi:2024ert}%
  \BibitemOpen
  \bibfield  {author} {\bibinfo {author} {\bibfnamefont {S.}~\bibnamefont
  {Pi}}, \bibinfo {author} {\bibfnamefont {M.}~\bibnamefont {Sasaki}}, \bibinfo
  {author} {\bibfnamefont {V.}~\bibnamefont {Takhistov}}, \ and\ \bibinfo
  {author} {\bibfnamefont {J.}~\bibnamefont {Wang}},\ }\href {\doibase
  10.1088/1475-7516/2025/09/045} {\bibfield  {journal} {\bibinfo  {journal}
  {JCAP}\ }\textbf {\bibinfo {volume} {09}},\ \bibinfo {pages} {045} (\bibinfo
  {year} {2025})},\ \Eprint {http://arxiv.org/abs/2501.00295} {arXiv:2501.00295
  [astro-ph.CO]} \BibitemShut {NoStop}%
\bibitem [{\citenamefont {Owen}(1980)}]{Owen1980}%
  \BibitemOpen
  \bibfield  {author} {\bibinfo {author} {\bibfnamefont {D.~B.}\ \bibnamefont
  {Owen}},\ }\href {\doibase 10.1080/03610918008812164} {\bibfield  {journal}
  {\bibinfo  {journal} {Communications in Statistics - Simulation and
  Computation}\ }\textbf {\bibinfo {volume} {9}},\ \bibinfo {pages} {389}
  (\bibinfo {year} {1980})}\BibitemShut {NoStop}%
\bibitem [{\citenamefont {Pattison}\ \emph {et~al.}(2017)\citenamefont
  {Pattison}, \citenamefont {Vennin}, \citenamefont {Assadullahi},\ and\
  \citenamefont {Wands}}]{Pattison:2017mbe}%
  \BibitemOpen
  \bibfield  {author} {\bibinfo {author} {\bibfnamefont {C.}~\bibnamefont
  {Pattison}}, \bibinfo {author} {\bibfnamefont {V.}~\bibnamefont {Vennin}},
  \bibinfo {author} {\bibfnamefont {H.}~\bibnamefont {Assadullahi}}, \ and\
  \bibinfo {author} {\bibfnamefont {D.}~\bibnamefont {Wands}},\ }\href
  {\doibase 10.1088/1475-7516/2017/10/046} {\bibfield  {journal} {\bibinfo
  {journal} {JCAP}\ }\textbf {\bibinfo {volume} {10}},\ \bibinfo {pages} {046}
  (\bibinfo {year} {2017})},\ \Eprint {http://arxiv.org/abs/1707.00537}
  {arXiv:1707.00537 [hep-th]} \BibitemShut {NoStop}%
\bibitem [{\citenamefont {Franciolini}\ \emph {et~al.}(2018)\citenamefont
  {Franciolini}, \citenamefont {Kehagias}, \citenamefont {Matarrese},\ and\
  \citenamefont {Riotto}}]{Franciolini:2018vbk}%
  \BibitemOpen
  \bibfield  {author} {\bibinfo {author} {\bibfnamefont {G.}~\bibnamefont
  {Franciolini}}, \bibinfo {author} {\bibfnamefont {A.}~\bibnamefont
  {Kehagias}}, \bibinfo {author} {\bibfnamefont {S.}~\bibnamefont {Matarrese}},
  \ and\ \bibinfo {author} {\bibfnamefont {A.}~\bibnamefont {Riotto}},\ }\href
  {\doibase 10.1088/1475-7516/2018/03/016} {\bibfield  {journal} {\bibinfo
  {journal} {JCAP}\ }\textbf {\bibinfo {volume} {03}},\ \bibinfo {pages} {016}
  (\bibinfo {year} {2018})},\ \Eprint {http://arxiv.org/abs/1801.09415}
  {arXiv:1801.09415 [astro-ph.CO]} \BibitemShut {NoStop}%
\bibitem [{\citenamefont {Ezquiaga}\ \emph {et~al.}(2020)\citenamefont
  {Ezquiaga}, \citenamefont {Garc{\'\i}a-Bellido},\ and\ \citenamefont
  {Vennin}}]{Ezquiaga:2019ftu}%
  \BibitemOpen
  \bibfield  {author} {\bibinfo {author} {\bibfnamefont {J.~M.}\ \bibnamefont
  {Ezquiaga}}, \bibinfo {author} {\bibfnamefont {J.}~\bibnamefont
  {Garc{\'\i}a-Bellido}}, \ and\ \bibinfo {author} {\bibfnamefont
  {V.}~\bibnamefont {Vennin}},\ }\href {\doibase 10.1088/1475-7516/2020/03/029}
  {\bibfield  {journal} {\bibinfo  {journal} {JCAP}\ }\textbf {\bibinfo
  {volume} {03}},\ \bibinfo {pages} {029} (\bibinfo {year} {2020})},\ \Eprint
  {http://arxiv.org/abs/1912.05399} {arXiv:1912.05399 [astro-ph.CO]}
  \BibitemShut {NoStop}%
\bibitem [{\citenamefont {Ando}\ and\ \citenamefont
  {Vennin}(2021)}]{Ando:2020fjm}%
  \BibitemOpen
  \bibfield  {author} {\bibinfo {author} {\bibfnamefont {K.}~\bibnamefont
  {Ando}}\ and\ \bibinfo {author} {\bibfnamefont {V.}~\bibnamefont {Vennin}},\
  }\href {\doibase 10.1088/1475-7516/2021/04/057} {\bibfield  {journal}
  {\bibinfo  {journal} {JCAP}\ }\textbf {\bibinfo {volume} {04}},\ \bibinfo
  {pages} {057} (\bibinfo {year} {2021})},\ \Eprint
  {http://arxiv.org/abs/2012.02031} {arXiv:2012.02031 [astro-ph.CO]}
  \BibitemShut {NoStop}%
\bibitem [{\citenamefont {Figueroa}\ \emph {et~al.}(2021)\citenamefont
  {Figueroa}, \citenamefont {Raatikainen}, \citenamefont {Rasanen},\ and\
  \citenamefont {Tomberg}}]{Figueroa:2020jkf}%
  \BibitemOpen
  \bibfield  {author} {\bibinfo {author} {\bibfnamefont {D.~G.}\ \bibnamefont
  {Figueroa}}, \bibinfo {author} {\bibfnamefont {S.}~\bibnamefont
  {Raatikainen}}, \bibinfo {author} {\bibfnamefont {S.}~\bibnamefont
  {Rasanen}}, \ and\ \bibinfo {author} {\bibfnamefont {E.}~\bibnamefont
  {Tomberg}},\ }\href {\doibase 10.1103/PhysRevLett.127.101302} {\bibfield
  {journal} {\bibinfo  {journal} {Phys. Rev. Lett.}\ }\textbf {\bibinfo
  {volume} {127}},\ \bibinfo {pages} {101302} (\bibinfo {year} {2021})},\
  \Eprint {http://arxiv.org/abs/2012.06551} {arXiv:2012.06551 [astro-ph.CO]}
  \BibitemShut {NoStop}%
\bibitem [{\citenamefont {Tada}\ and\ \citenamefont
  {Vennin}(2022)}]{Tada:2021zzj}%
  \BibitemOpen
  \bibfield  {author} {\bibinfo {author} {\bibfnamefont {Y.}~\bibnamefont
  {Tada}}\ and\ \bibinfo {author} {\bibfnamefont {V.}~\bibnamefont {Vennin}},\
  }\href {\doibase 10.1088/1475-7516/2022/02/021} {\bibfield  {journal}
  {\bibinfo  {journal} {JCAP}\ }\textbf {\bibinfo {volume} {02}},\ \bibinfo
  {pages} {021} (\bibinfo {year} {2022})},\ \Eprint
  {http://arxiv.org/abs/2111.15280} {arXiv:2111.15280 [astro-ph.CO]}
  \BibitemShut {NoStop}%
\bibitem [{\citenamefont {Kitajima}\ \emph {et~al.}(2021)\citenamefont
  {Kitajima}, \citenamefont {Tada}, \citenamefont {Yokoyama},\ and\
  \citenamefont {Yoo}}]{Kitajima:2021fpq}%
  \BibitemOpen
  \bibfield  {author} {\bibinfo {author} {\bibfnamefont {N.}~\bibnamefont
  {Kitajima}}, \bibinfo {author} {\bibfnamefont {Y.}~\bibnamefont {Tada}},
  \bibinfo {author} {\bibfnamefont {S.}~\bibnamefont {Yokoyama}}, \ and\
  \bibinfo {author} {\bibfnamefont {C.-M.}\ \bibnamefont {Yoo}},\ }\href
  {\doibase 10.1088/1475-7516/2021/10/053} {\bibfield  {journal} {\bibinfo
  {journal} {JCAP}\ }\textbf {\bibinfo {volume} {10}},\ \bibinfo {pages} {053}
  (\bibinfo {year} {2021})},\ \Eprint {http://arxiv.org/abs/2109.00791}
  {arXiv:2109.00791 [astro-ph.CO]} \BibitemShut {NoStop}%
\bibitem [{\citenamefont {Hooshangi}\ \emph {et~al.}(2022)\citenamefont
  {Hooshangi}, \citenamefont {Namjoo},\ and\ \citenamefont
  {Noorbala}}]{Hooshangi:2021ubn}%
  \BibitemOpen
  \bibfield  {author} {\bibinfo {author} {\bibfnamefont {S.}~\bibnamefont
  {Hooshangi}}, \bibinfo {author} {\bibfnamefont {M.~H.}\ \bibnamefont
  {Namjoo}}, \ and\ \bibinfo {author} {\bibfnamefont {M.}~\bibnamefont
  {Noorbala}},\ }\href {\doibase 10.1016/j.physletb.2022.137400} {\bibfield
  {journal} {\bibinfo  {journal} {Phys. Lett. B}\ }\textbf {\bibinfo {volume}
  {834}},\ \bibinfo {pages} {137400} (\bibinfo {year} {2022})},\ \Eprint
  {http://arxiv.org/abs/2112.04520} {arXiv:2112.04520 [astro-ph.CO]}
  \BibitemShut {NoStop}%
\bibitem [{\citenamefont {Biagetti}\ \emph {et~al.}(2021)\citenamefont
  {Biagetti}, \citenamefont {De~Luca}, \citenamefont {Franciolini},
  \citenamefont {Kehagias},\ and\ \citenamefont {Riotto}}]{Biagetti:2021eep}%
  \BibitemOpen
  \bibfield  {author} {\bibinfo {author} {\bibfnamefont {M.}~\bibnamefont
  {Biagetti}}, \bibinfo {author} {\bibfnamefont {V.}~\bibnamefont {De~Luca}},
  \bibinfo {author} {\bibfnamefont {G.}~\bibnamefont {Franciolini}}, \bibinfo
  {author} {\bibfnamefont {A.}~\bibnamefont {Kehagias}}, \ and\ \bibinfo
  {author} {\bibfnamefont {A.}~\bibnamefont {Riotto}},\ }\href {\doibase
  10.1016/j.physletb.2021.136602} {\bibfield  {journal} {\bibinfo  {journal}
  {Phys. Lett. B}\ }\textbf {\bibinfo {volume} {820}},\ \bibinfo {pages}
  {136602} (\bibinfo {year} {2021})},\ \Eprint
  {http://arxiv.org/abs/2105.07810} {arXiv:2105.07810 [astro-ph.CO]}
  \BibitemShut {NoStop}%
\bibitem [{\citenamefont {Gow}\ \emph {et~al.}(2023)\citenamefont {Gow},
  \citenamefont {Assadullahi}, \citenamefont {Jackson}, \citenamefont {Koyama},
  \citenamefont {Vennin},\ and\ \citenamefont {Wands}}]{Gow:2022jfb}%
  \BibitemOpen
  \bibfield  {author} {\bibinfo {author} {\bibfnamefont {A.~D.}\ \bibnamefont
  {Gow}}, \bibinfo {author} {\bibfnamefont {H.}~\bibnamefont {Assadullahi}},
  \bibinfo {author} {\bibfnamefont {J.~H.~P.}\ \bibnamefont {Jackson}},
  \bibinfo {author} {\bibfnamefont {K.}~\bibnamefont {Koyama}}, \bibinfo
  {author} {\bibfnamefont {V.}~\bibnamefont {Vennin}}, \ and\ \bibinfo {author}
  {\bibfnamefont {D.}~\bibnamefont {Wands}},\ }\href {\doibase
  10.1209/0295-5075/acd417} {\bibfield  {journal} {\bibinfo  {journal} {EPL}\
  }\textbf {\bibinfo {volume} {142}},\ \bibinfo {pages} {49001} (\bibinfo
  {year} {2023})},\ \Eprint {http://arxiv.org/abs/2211.08348} {arXiv:2211.08348
  [astro-ph.CO]} \BibitemShut {NoStop}%
\bibitem [{\citenamefont {Raatikainen}\ \emph {et~al.}(2024)\citenamefont
  {Raatikainen}, \citenamefont {R{\"a}s{\"a}nen},\ and\ \citenamefont
  {Tomberg}}]{Raatikainen:2023bzk}%
  \BibitemOpen
  \bibfield  {author} {\bibinfo {author} {\bibfnamefont {S.}~\bibnamefont
  {Raatikainen}}, \bibinfo {author} {\bibfnamefont {S.}~\bibnamefont
  {R{\"a}s{\"a}nen}}, \ and\ \bibinfo {author} {\bibfnamefont {E.}~\bibnamefont
  {Tomberg}},\ }\href {\doibase 10.1103/PhysRevLett.133.121403} {\bibfield
  {journal} {\bibinfo  {journal} {Phys. Rev. Lett.}\ }\textbf {\bibinfo
  {volume} {133}},\ \bibinfo {pages} {121403} (\bibinfo {year} {2024})},\
  \Eprint {http://arxiv.org/abs/2312.12911} {arXiv:2312.12911 [astro-ph.CO]}
  \BibitemShut {NoStop}%
\bibitem [{\citenamefont {Firouzjahi}\ and\ \citenamefont
  {Riotto}(2023)}]{Firouzjahi:2023xke}%
  \BibitemOpen
  \bibfield  {author} {\bibinfo {author} {\bibfnamefont {H.}~\bibnamefont
  {Firouzjahi}}\ and\ \bibinfo {author} {\bibfnamefont {A.}~\bibnamefont
  {Riotto}},\ }\href {\doibase 10.1103/PhysRevD.108.123504} {\bibfield
  {journal} {\bibinfo  {journal} {Phys. Rev. D}\ }\textbf {\bibinfo {volume}
  {108}},\ \bibinfo {pages} {123504} (\bibinfo {year} {2023})},\ \Eprint
  {http://arxiv.org/abs/2309.10536} {arXiv:2309.10536 [astro-ph.CO]}
  \BibitemShut {NoStop}%
\bibitem [{\citenamefont {Animali}\ and\ \citenamefont
  {Vennin}(2024)}]{Animali:2024jiz}%
  \BibitemOpen
  \bibfield  {author} {\bibinfo {author} {\bibfnamefont {C.}~\bibnamefont
  {Animali}}\ and\ \bibinfo {author} {\bibfnamefont {V.}~\bibnamefont
  {Vennin}},\ }\href {\doibase 10.1088/1475-7516/2024/08/026} {\bibfield
  {journal} {\bibinfo  {journal} {JCAP}\ }\textbf {\bibinfo {volume} {08}},\
  \bibinfo {pages} {026} (\bibinfo {year} {2024})},\ \Eprint
  {http://arxiv.org/abs/2402.08642} {arXiv:2402.08642 [astro-ph.CO]}
  \BibitemShut {NoStop}%
\bibitem [{\citenamefont {Vennin}\ and\ \citenamefont
  {Wands}(2025)}]{Vennin:2024yzl}%
  \BibitemOpen
  \bibfield  {author} {\bibinfo {author} {\bibfnamefont {V.}~\bibnamefont
  {Vennin}}\ and\ \bibinfo {author} {\bibfnamefont {D.}~\bibnamefont {Wands}},\
  }\enquote {\bibinfo {title} {{Quantum Diffusion and~Large Primordial
  Perturbations from~Inflation}},}\ \ (\bibinfo {year} {2025})\ \Eprint
  {http://arxiv.org/abs/2402.12672} {arXiv:2402.12672 [astro-ph.CO]}
  \BibitemShut {NoStop}%
\bibitem [{\citenamefont {Jackson}\ \emph {et~al.}(2025)\citenamefont
  {Jackson}, \citenamefont {Assadullahi}, \citenamefont {Gow}, \citenamefont
  {Koyama}, \citenamefont {Vennin},\ and\ \citenamefont
  {Wands}}]{Jackson:2024aoo}%
  \BibitemOpen
  \bibfield  {author} {\bibinfo {author} {\bibfnamefont {J.~H.~P.}\
  \bibnamefont {Jackson}}, \bibinfo {author} {\bibfnamefont {H.}~\bibnamefont
  {Assadullahi}}, \bibinfo {author} {\bibfnamefont {A.~D.}\ \bibnamefont
  {Gow}}, \bibinfo {author} {\bibfnamefont {K.}~\bibnamefont {Koyama}},
  \bibinfo {author} {\bibfnamefont {V.}~\bibnamefont {Vennin}}, \ and\ \bibinfo
  {author} {\bibfnamefont {D.}~\bibnamefont {Wands}},\ }\href {\doibase
  10.1088/1475-7516/2025/04/073} {\bibfield  {journal} {\bibinfo  {journal}
  {JCAP}\ }\textbf {\bibinfo {volume} {04}},\ \bibinfo {pages} {073} (\bibinfo
  {year} {2025})},\ \Eprint {http://arxiv.org/abs/2410.13683} {arXiv:2410.13683
  [astro-ph.CO]} \BibitemShut {NoStop}%
\bibitem [{\citenamefont {Ianniccari}\ \emph {et~al.}(2024)\citenamefont
  {Ianniccari}, \citenamefont {Iovino}, \citenamefont {Kehagias}, \citenamefont
  {Perrone},\ and\ \citenamefont {Riotto}}]{Ianniccari:2024bkh}%
  \BibitemOpen
  \bibfield  {author} {\bibinfo {author} {\bibfnamefont {A.}~\bibnamefont
  {Ianniccari}}, \bibinfo {author} {\bibfnamefont {A.~J.}\ \bibnamefont
  {Iovino}}, \bibinfo {author} {\bibfnamefont {A.}~\bibnamefont {Kehagias}},
  \bibinfo {author} {\bibfnamefont {D.}~\bibnamefont {Perrone}}, \ and\
  \bibinfo {author} {\bibfnamefont {A.}~\bibnamefont {Riotto}},\ }\href
  {\doibase 10.1103/PhysRevD.109.123549} {\bibfield  {journal} {\bibinfo
  {journal} {Phys. Rev. D}\ }\textbf {\bibinfo {volume} {109}},\ \bibinfo
  {pages} {123549} (\bibinfo {year} {2024})},\ \Eprint
  {http://arxiv.org/abs/2402.11033} {arXiv:2402.11033 [astro-ph.CO]}
  \BibitemShut {NoStop}%
\bibitem [{\citenamefont {Animali}\ \emph {et~al.}(2025)\citenamefont
  {Animali}, \citenamefont {Auclair}, \citenamefont {Blachier},\ and\
  \citenamefont {Vennin}}]{Animali:2025pyf}%
  \BibitemOpen
  \bibfield  {author} {\bibinfo {author} {\bibfnamefont {C.}~\bibnamefont
  {Animali}}, \bibinfo {author} {\bibfnamefont {P.}~\bibnamefont {Auclair}},
  \bibinfo {author} {\bibfnamefont {B.}~\bibnamefont {Blachier}}, \ and\
  \bibinfo {author} {\bibfnamefont {V.}~\bibnamefont {Vennin}},\ }\href
  {\doibase 10.1088/1475-7516/2025/05/019} {\bibfield  {journal} {\bibinfo
  {journal} {JCAP}\ }\textbf {\bibinfo {volume} {05}},\ \bibinfo {pages} {019}
  (\bibinfo {year} {2025})},\ \Eprint {http://arxiv.org/abs/2501.05371}
  {arXiv:2501.05371 [astro-ph.CO]} \BibitemShut {NoStop}%
\bibitem [{\citenamefont {{Auclair}}(2025)}]{2025zndo..15235932A}%
  \BibitemOpen
  \bibfield  {author} {\bibinfo {author} {\bibfnamefont {P.}~\bibnamefont
  {{Auclair}}},\ }\href {\doibase 10.5281/zenodo.15235932} {\enquote {\bibinfo
  {title} {{FOREST: FOrtran Recursive Exploration of Stochastic Trees}},}\ }
  (\bibinfo {year} {2025})\BibitemShut {NoStop}%
\bibitem [{\citenamefont {Choudhury}(2025)}]{Choudhury:2025kxg}%
  \BibitemOpen
  \bibfield  {author} {\bibinfo {author} {\bibfnamefont {S.}~\bibnamefont
  {Choudhury}},\ }\href {\doibase 10.1142/S0218271825440237} {\bibfield
  {journal} {\bibinfo  {journal} {Int. J. Mod. Phys. D}\ }\textbf {\bibinfo
  {volume} {34}},\ \bibinfo {pages} {2544023} (\bibinfo {year} {2025})},\
  \Eprint {http://arxiv.org/abs/2503.17635} {arXiv:2503.17635 [gr-qc]}
  \BibitemShut {NoStop}%
\end{thebibliography}%

\end{document}